\documentclass[10pt,twocolumn,letterpaper]{article}

\usepackage{cvpr}
\usepackage{times}
\usepackage{epsfig}
\usepackage{graphicx}
\usepackage{amsmath}
\usepackage{amssymb}
\usepackage{array}
\usepackage{subfigure}
\usepackage{multirow}
\usepackage{tabularx}
\usepackage{makecell}
\usepackage{adjustbox}
\usepackage[numbers, sort]{natbib}

\newcolumntype{L}[1]{>{\raggedright\arraybackslash}p{#1}}
\newcolumntype{C}[1]{>{\centering\arraybackslash}p{#1}}
\newcolumntype{R}[1]{>{\raggedleft\arraybackslash}p{#1}}

\newcommand{\widthscalefive}{0.16}

\newcommand{\picwidth}{0.24}
\newcommand{\pagehspace}{-1pt}
\newcommand{\pagevspace}{2pt}
\newcommand{\picvspace}{-3pt}
\newcommand{\captionvspace}{0pt}
\newcommand{\packvspace}{6pt}

\usepackage[pagebackref=true,breaklinks=true,letterpaper=true,colorlinks,bookmarks=false]{hyperref}

\cvprfinalcopy 

\def\cvprPaperID{2888} 

\ifcvprfinal\fi
\begin{document}

\title{Structure-Preserving Super Resolution with Gradient Guidance}

\author{
Cheng Ma$^{1,2,3}$, Yongming Rao$^{1,2,3}$, Yean Cheng$^{1}$, Ce Chen$^{1}$, Jiwen Lu$^{1,2,3}\thanks{Corresponding author}$\ , Jie Zhou$^{1,2,3,4}$\\
{$^1$Department of Automation, Tsinghua University, China}\\
{$^2$State Key Lab of Intelligent Technologies and Systems, China}\\
{$^3$Beijing National Research Center for Information Science and Technology, China}\\
{$^4$Tsinghua Shenzhen International Graduate School, Tsinghua University, China} \\
{\tt\small macheng17@mails.tsinghua.edu.cn; raoyongming95@gmail.com} \\
{\tt\small \{cya17, chence17\}@mails.tsinghua.edu.cn; \{lujiwen, jzhou\}@tsinghua.edu.cn}\\
}

\maketitle
\thispagestyle{empty}

\begin{abstract}
   Structures matter in single image super resolution (SISR).
   Recent studies benefiting from generative adversarial network (GAN) have promoted the development of SISR by recovering photo-realistic images. However, there are always undesired structural distortions in the recovered images. 
   In this paper, we propose a structure-preserving super resolution method to alleviate the above issue while maintaining the merits of GAN-based methods to generate perceptual-pleasant details. Specifically, we exploit gradient maps of images to guide the recovery in two aspects. 
   On the one hand, we restore high-resolution gradient maps by a gradient branch to provide additional structure priors for the SR process. 
   On the other hand, we propose a gradient loss which imposes a second-order restriction on the super-resolved images. Along with the previous image-space loss functions, the gradient-space objectives help generative networks concentrate more on geometric structures. Moreover, our method is model-agnostic, which can be potentially used for off-the-shelf SR networks.
   Experimental results show that we achieve the best PI and LPIPS performance and meanwhile comparable PSNR and SSIM compared with state-of-the-art perceptual-driven SR methods. Visual results demonstrate our superiority in restoring structures while generating natural SR images. \footnote{Code: \href{https://github.com/Maclory/SPSR}{https://github.com/Maclory/SPSR}}
\end{abstract}

\section{Introduction}

\begin{figure}[htbp]
\centering

\subfigure[HR]{
\begin{minipage}[b]{0.48\linewidth}
\includegraphics[width=1 \linewidth]{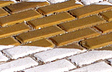}
\end{minipage}%
}%
\subfigure[RCAN~\cite{RCAN}]{
\begin{minipage}[b]{0.48\linewidth}
\includegraphics[width=1 \linewidth]{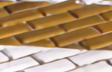}
\end{minipage}
}
\vspace{-3mm}

\subfigure[SRGAN~\cite{ledig2017photo}]{
\begin{minipage}[b]{0.48\linewidth}
\includegraphics[width=1 \linewidth]{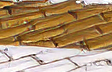}
\end{minipage}%
}%
\subfigure[ESRGAN~\cite{wang2018esrgan}]{
\begin{minipage}[b]{0.48\linewidth}
\includegraphics[width=1 \linewidth]{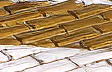}
\end{minipage}
}
\vspace{-3mm}

\subfigure[NatSR~\cite{soh2019natural}]{
\begin{minipage}[b]{0.48\linewidth}
\includegraphics[width=1 \linewidth]{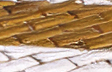}
\end{minipage}%
}%
\subfigure[SPSR (Ours)]{
\begin{minipage}[b]{0.48\linewidth}
\includegraphics[width=1 \linewidth]{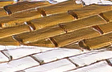}
\end{minipage}
}

\centering
\caption{SR results of different methods. RCAN represents PSNR-oriented methods, typically generating straight but blurry edges for the bricks. Perceptual-driven methods including SRGAN, ESRGAN and NatSR commonly recover sharper but geometric-inconsistent textures. Our SPSR result is sharper than that of RCAN, and preserve finer geometric structures compared with perceptual-driven methods. Best viewed on screen. }
\vspace{-2mm}
\label{fig:head}
\end{figure}

Single image super resolution (SISR) aims to recover high-resolution (HR) images from their low-resolution (LR) counterparts. SISR is a fundamental problem in the community of computer vision and can be applied in many image analysis tasks including surveillance and satellite image. It is a widely known ill-posed problem since each LR input may have multiple HR solutions. With the development of deep learning, a number of SR methods~\cite{dong2014learning,shi2016real} have been proposed. Most of them are optimized by the mean squared error (MSE) which measures the pixel-wise distances between SR images and the HR ones. However, such optimizing objective impels a deep model to produce an image which may be a statistical average of possible HR solutions to the one-to-many problem. As a result, such methods usually generate blurry images with high peak signal-to-noise ratio (PSNR). 

Hence, several methods aiming to recover photo-realistic images have recently utilized the generative adversarial network (GAN)~\cite{goodfellow2014generative}, such as SRGAN~\cite{ledig2017photo}, EnhanceNet~\cite{EnhanceNet}, ESRGAN~\cite{wang2018esrgan} and NatSR~\cite{soh2019natural}. 
While GAN-based methods can generate high-fidelity SR results, there are always geometric distortions along with sharp edges and fine textures. Some SR examples are presented in Figure~\ref{fig:head}. We can see RCAN~\cite{RCAN} recovers blurry but straight edges for the bricks, while edges restored by perceptual-driven methods are sharper but twisted. 
In fact, GAN-based methods generally suffer from structural inconsistency since the discriminators may introduce unstable factors to the optimization procedure. Some methods have been proposed to balance the trade-off between the merits of two kinds of SR methods. For example, Controllable Feature Space Network (CFSNet)~\cite{wang2019cfsnet} designs an interactive framework to transfer continuously between two objectives of perceptual quality and distortion reduction. Nevertheless, the intrinsic problem is not mitigated since the two goals cannot be achieved simultaneously. Hence it is necessary to explicitly guide perceptual-driven SR methods to preserve structures for further enhancing the SR performance. 

In this paper, we propose a structure-preserving super resolution method to alleviate the above-mentioned issue. Since the gradient map reveals the sharpness of each local region in an image, we exploit this powerful tool to guide image recovery. 
On the one hand, we design a gradient branch which converts the gradient maps of  LR images to the HR ones as an auxiliary SR problem. The recovered gradients can be integrated into the SR branch to provide structure prior for SR. Besides, the gradients can highlight the regions where sharpness and structures should be paid more attention to, so as to guide the high-quality generation explicitly. 
This idea is motivated by the observation that once edges are recovered with high-fidelity, the SR task can be treated as a color-filling problem with strong clues given by the LR images. 
On the other hand, we propose a gradient loss to explicitly supervise the gradient maps of recovered images. Together with the image-space loss functions in existing methods, the gradient loss restricts the second-order relationship of neighboring pixels. Hence the structural configuration can be better retained with such guidance, and  the SR results with high perceptual quality and fewer geometric distortions can be obtained. Moreover, our method is model-agnostic, which can be potentially used for off-the-shelf SR networks. To the best of our knowledge, we are the first to explicitly consider preserving geometric structures in GAN-based SR methods. 
Experimental results on benchmark datasets show that our method succeeds in enhancing SR fidelity by reducing structural distortions.

\section{Related Work}

Here we review SISR methods~\cite{classicalCubic,classicalExampleBased,sun2008image,duchon1979lanczos,classicalNeighborEmbedding,irani1991improving,xiong2010robust,glasner2009super,freedman2011image,kim2010single,yang2010image,yang2008image} which can be classified into two categories: PSNR-oriented methods and perceptual-driven ones. We also investigate methods relevant to gradient.

\textbf{PSNR-Oriented Methods}: Most previous approaches target high PSNR. As a pioneer, Dong \etal~\cite{dong2014learning} propose SRCNN, which firstly maps LR images to HR ones by a three-layer CNN. DRCN~\cite{DRCN} and VDSR~\cite{VDSR} are further proposed by Kim \etal to improve SR performance. 
Moreover, Ledig \etal~\cite{ledig2017photo} propose SRResNet by employing the idea of ResNet~\cite{he2016deep}. 
Zhang \etal~\cite{RDN} propose RDN by utilizing residual dense blocks in the SR framework. They further introduce RCAN~\cite{RCAN}
and achieve superior performance on PSNR. Li \etal~\cite{li2019feedback} propose a feedback framework to refine the super-resolved results step by step. 

\begin{figure*}
\begin{center}
\includegraphics[width=\linewidth]{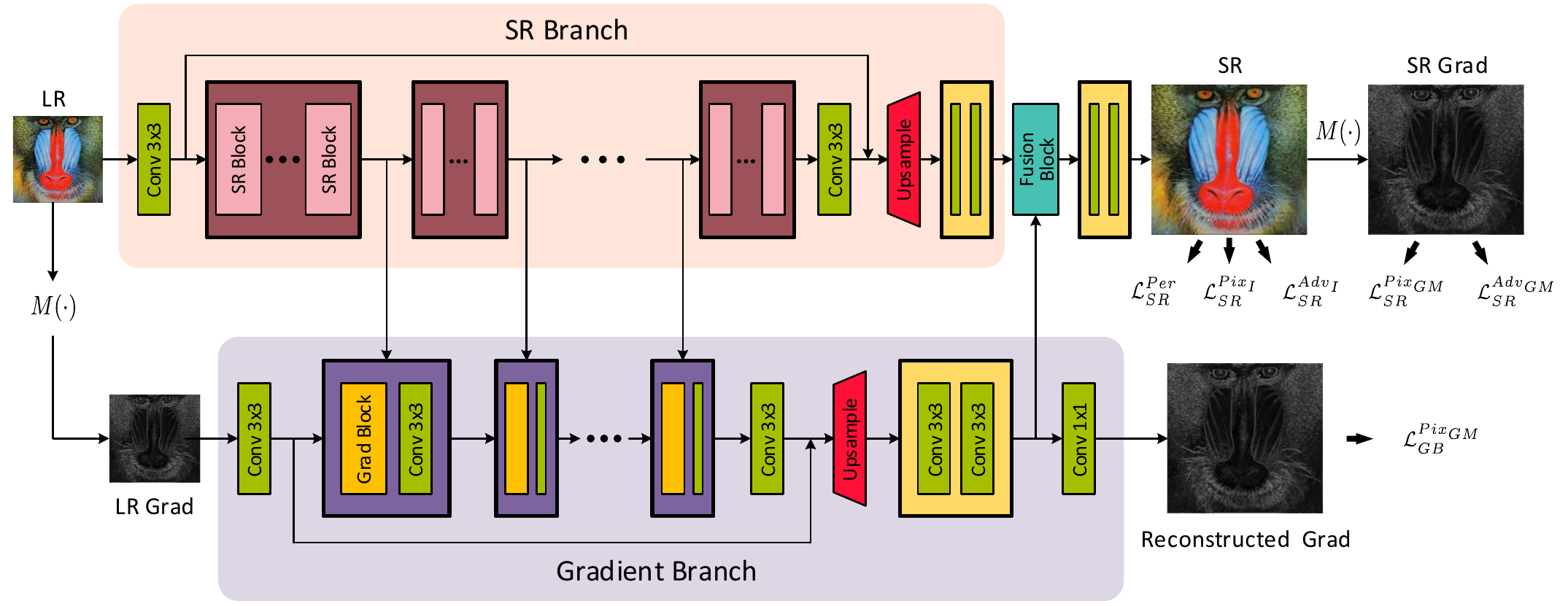}
\vspace{-4mm}
\end{center}
   \caption{Overall framework of our SPSR method. Our architecture consists of two branches, the SR branch and the gradient branch. The gradient branch aims to super-resolve LR gradient maps to the HR counterparts. It incorporates multi-level representations from the SR branch to reduce parameters and outputs gradient information to guide the SR process by a fusion block in turn. The final SR outputs are optimized by not only conventional image-space losses, but also the proposed gradient-space objectives. }
\label{fig:framework}
\vspace{-2mm}
\end{figure*}

\textbf{Perceptual-Driven Methods}: The methods mentioned above all focus on achieving high PSNR and thus use the MSE loss or L1 loss as loss functions. 
However, these methods usually produce blurry images.
Johnson \etal~\cite{johnson2016perceptual} propose perceptual loss to improve the visual quality of recovered images. 
Ledig \etal~\cite{ledig2017photo} utilize adversarial loss~\cite{goodfellow2014generative} to construct SRGAN, which becomes the first framework able to generate photo-realistic HR images. Furthermore, Sajjadi \etal~\cite{EnhanceNet} restore high-fidelity textures by texture loss. Wang \etal~\cite{wang2018esrgan} enhance the previous frameworks by introducing Residual-in-Residual Dense Block (RRDB) to the proposed ESRGAN. 
Wang \etal~\cite{SFTGAN} 
exploit semantic segmentation maps as priors to generate more natural textures for specific categories. Rad \etal~\cite{rad2019srobb} propose a targeted perceptual loss on the basis of the labels of object, background and boundary. 
Although these existing perceptual-driven methods indeed improve the overall visual quality of super-resolved images, they sometimes generate unnatural artifacts including geometric distortions when recovering details.

\textbf{Gradient-Relevant Methods}: Gradient information has been utilized in previous work~\cite{luan2017deep,anoosheh2019night}.
For SR methods, Fattal~\cite{fattal2007image} proposes a method based on edge statistics of image gradients by learning
the prior dependency of different resolutions. 
Sun \etal~\cite{sun2010gradient} propose a gradient profile prior to represent image gradients and a gradient field transformation to enhance sharpness of super-resolved images.  
Yan \etal~\cite{yan2015single} propose a SR method based on gradient profile sharpness which is extracted from gradient description models. 
In these methods, statistical dependencies are modeled by estimating HR edge-related parameters according to those observed in LR images. However, the modeling procedure is accomplished point by point, which is complex and inflexible.  
In fact, deep learning is outstanding in handling probability transformation over the distribution of pixels. However, few methods have utilized its powerful abilities in gradient-relevant SR methods. 
Moreover, Zhu \etal~\cite{zhu2015modeling} propose a gradient-based SR method by collecting a dictionary of gradient patterns and modeling deformable gradient compositions. Yang \etal~\cite{yang2017deep} propose a recurrent residual network to reconstruct fine details guided by the edges which are extracted by off-the-shelf edge detector. 
While edge reconstruction and gradient field constraint have been utilized in some methods, their purposes are mainly to recover high-frequency components for PSNR-orientated SR methods. 
Different from these methods, we aim to reduce geometric distortions produced by GAN-based methods and exploit gradient maps as structure guidance for SR. For deep adversarial networks, gradient-space constraint may provide additional supervision for better image reconstruction. 
To the best of our knowledge, no GAN-based SR method has exploited gradient-space guidance for preserving texture structures. In this work, we aim to leverage gradient information to further improve the GAN-based SR methods.

\section{Approach}

In this section, we first introduce the overall framework. Then we present the details of gradient branch, attentive fusion module and final objective functions accordingly. 

\subsection{Overview}

In SISR, we aim to take LR images $I^{LR}$ as inputs and generate SR images $I^{SR}$ given their HR counterparts $I^{HR}$ as ground-truth. We denote the generator as $G$ and its parameters as $\theta_G$ and then we have $I^{SR} = G(I^{LR}; \theta_G)$. $I^{SR}$ should be as similar to $I^{HR}$ as possible. If the parameters are optimized by an loss function $\mathcal{L}$, we have the following formulation: 
\begin{eqnarray}
\theta_G^* &=& \arg \mathop{\min}_{\theta_G} \mathbb{E}_{I^{SR}}\mathcal{L}(G(I^{LR}; \theta_G), I^{HR}). 
\end{eqnarray}

The overall framework is depicted as Figure~\ref{fig:framework}. The generator is composed of two branches, one of which is a structure-preserving SR branch and the other is a gradient branch. The SR branch takes $I^{LR}$ as input and aims to recover the SR output $I^{SR}$ with the guidance provided by the SR gradient map from the gradient branch. 

\subsection{Details in Architecture}

\subsubsection{Gradient Branch}

The target of the gradient branch is to estimate the translation of gradient maps from the LR modality to the HR one. The gradient map for an image $I$ is obtained by computing the difference between adjacent pixels:
\begin{eqnarray}
I_x(\mathbf{x}) &=& I(x+1,y)-I(x-1, y), \nonumber \\
I_y(\mathbf{x}) &=& I(x, y+1)-I(x, y-1), \nonumber \\
\nabla I(\mathbf{x}) &=& (I_x(\mathbf{x}), I_y(\mathbf{x})), \nonumber \\
M(I) &=& \|\nabla I\|_2,
\end{eqnarray}
where $M(\cdot)$ stands for the operation to extract gradient map whose elements are gradient lengths for pixels with coordinates $\mathbf{x}=(x,y)$. The operation to get the gradients can be easily achieved by a convolution layer with a fixed kernel. In fact, we do not consider gradient direction information since gradient intensity is adequate to reveal the sharpness of local regions in recovered images. Hence we adopt the intensity maps as the gradient maps. 
Such gradient maps can be regarded as another kind of images, so that techniques for image-to-image translation can be utilized to learn the mapping between two modalities. The translation process is equivalent to the spatial distribution translation from LR edge sharpness to HR edge sharpness. Since most area of the gradient map is close to zero, the convolutional neural network can concentrates more on the spatial relationship of outlines. Therefore, it may be easier for the network to capture structure dependency and consequently produce approximate gradient maps for SR images. 

As shown in Figure~\ref{fig:framework}, the gradient branch incorporates several intermediate-level representations from the SR branch. The motivation of such scheme is that the well-designed SR branch is capable of carrying rich structural information which is pivotal to the recovery of gradient maps. Hence we utilize the features as a strong prior to promote the performance of the gradient branch, whose parameters can be largely reduced in this case. Between each two intermediate features, there is a gradient block which can be any basic block to extract higher-level features. 
Once we get the SR gradient maps by the gradient branch, we are able to integrate the obtained gradient features into the SR branch to guide SR reconstruction in turn. The magnitude of gradient map can implicitly reflect whether a recovered region should be sharp or smooth. In practice, we feed the feature maps produced by the next-to-last layer of gradient branch to the SR branch. Meanwhile, we generate the output gradient maps by a $1\times1$ convolution layer with these feature maps as inputs. 

\subsubsection{Structure-Preserving SR Branch}

We design a structure-preserving SR branch to get the final SR outputs. This branch constitutes of two parts. The first part is a regular SR network comprising of multiple generative neural blocks which can be any architecture. Here we introduce the Residual in Residual Dense Block (RRDB) proposed in ESRGAN~\cite{wang2018esrgan}. There are 23 RRDB blocks in the original model. Therefore, we incorporate the feature maps from the 5th, 10th, 15th, 20th blocks to the gradient branch. Since regular SR models produce images with only 3 channels, we remove the last convolutional reconstruction layer and feed the output feature to the consecutive part. 
The second part of the SR branch wires the SR gradient feature maps obtained from the gradient branch as mentioned above. We fuse the structure information by a fusion block which fuses the features from two branches together. Specifically, we concatenate the two features and then use another RRDB block and convolutional layer to reconstruct the final SR features. It is noteworthy that we only add one RRDB block into the SR branch. Thus the parameter increment is slight compared to the original model with 23 blocks. 

\subsection{Objective Functions}

\textbf{Conventional Loss}: Most SR methods optimize the elaborately designed networks by a common pixelwise loss, which is efficient for the task of super resolution measured by PSNR. This metric can reduce the average pixel difference between recovered images and ground-truths but the results may be too smooth to maintain sharp edges for visual effects. However, this loss is still widely used to accelerate convergence and improve SR performance: 
\begin{eqnarray}
\mathcal{L}^{Pix_I}_{SR} &=& \mathbb{E}_{I^{SR}} \|G(I^{LR})-I^{HR}\|_1.
\end{eqnarray}

Perceptual loss has been proposed in~\cite{johnson2016perceptual} to improve perceptual quality of recovered images. Features containing semantic information are extracted by a pre-trained VGG network~\cite{simonyan2014very}. The Euclidean distances between the features of HR images and SR ones are minimized in perceptual loss:
\begin{eqnarray}
\mathcal{L}^{Per}_{SR} &=& \mathbb{E}_{I^{SR}} \|\phi_i(G(I^{LR}))-\phi_i(I^{HR})\|_1, 
\end{eqnarray}
where $\phi_i(.)$ denotes the $i$th layer output of the VGG model.

Methods~\cite{ledig2017photo,wang2018esrgan} based on generative adversarial networks (GANs)~\cite{goodfellow2014generative,RaGAN,radford2015unsupervised,arjovsky2017wasserstein,gulrajani2017improved,berthelot2017began} also play an important role in the SR problem. The discriminator $D_I$ and the generator $G$ are optimized by a two-player game as follows: 
\begin{eqnarray}
\mathcal{L}^{{Dis}_I}_{SR} &=&  -\mathbb{E}_{I^{SR}}[\log (1-D_I(I^{SR}))] \nonumber \\
&&-\mathbb{E}_{I^{HR}}[\log D_I(I^{HR})], \\
\mathcal{L}^{{Adv}_I}_{SR} &=& -\mathbb{E}_{I^{SR}}[\log D_I(G(I^{LR}))].
\end{eqnarray}

Following~\cite{RaGAN,wang2018esrgan} we conduct relativistic average GAN (RaGAN) to achieve better optimization in practice. Models supervised by the above objective functions merely consider the image-space constraint for images, but neglect the semantically structural information provided by the gradient space. While the generated results look  photo-realistic, there are also a number of undesired geometric distortions. Thus we introduce the gradient loss to alleviate this issue. 

\begin{figure}[t]
\centering

\subfigure[HR]{
\begin{minipage}[b]{0.3\linewidth}
\includegraphics[width=1 \linewidth]{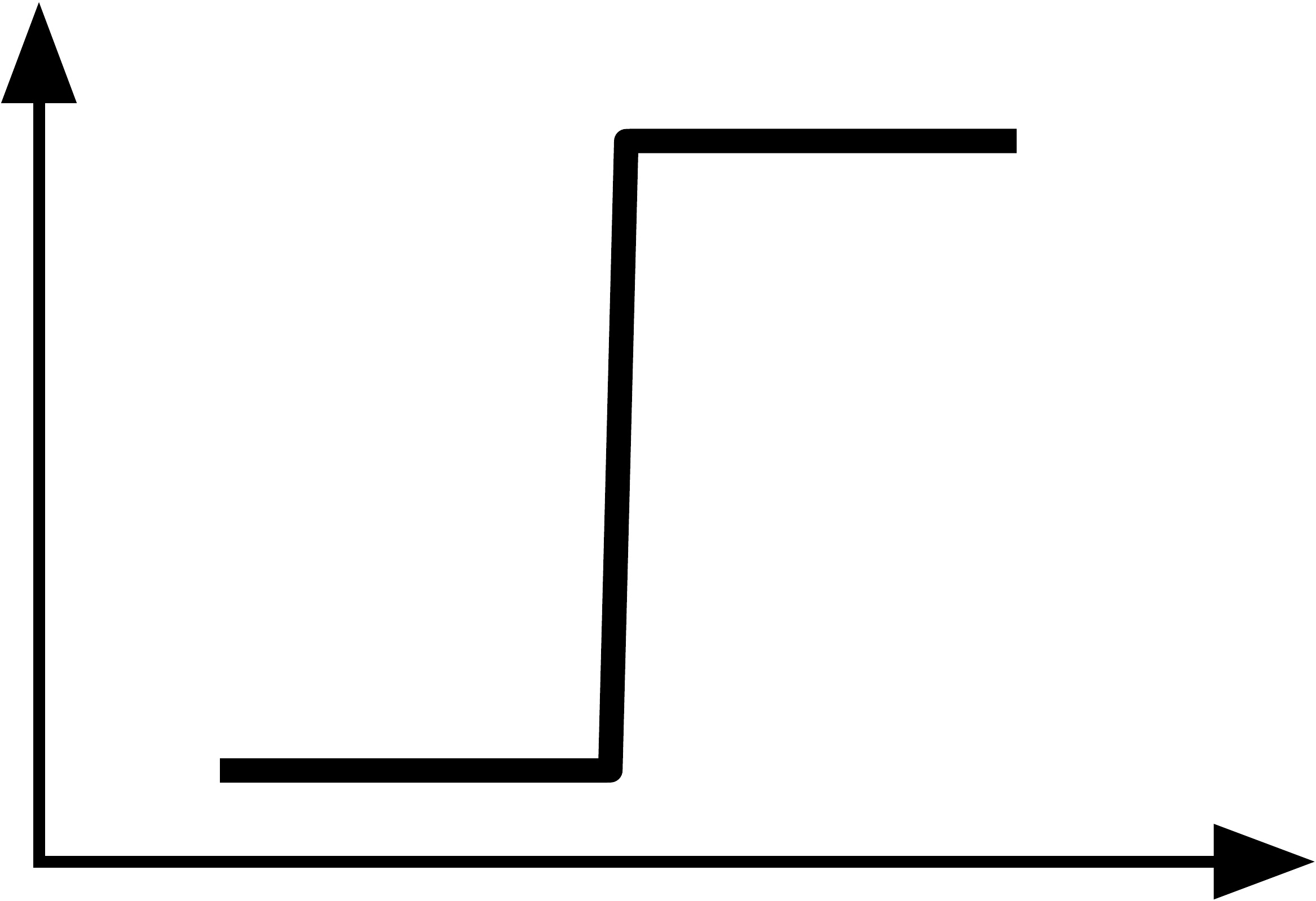}
\end{minipage}%
}\hspace{1pt}
\subfigure[Blurry SR]{
\begin{minipage}[b]{0.3\linewidth}
\includegraphics[width=1 \linewidth]{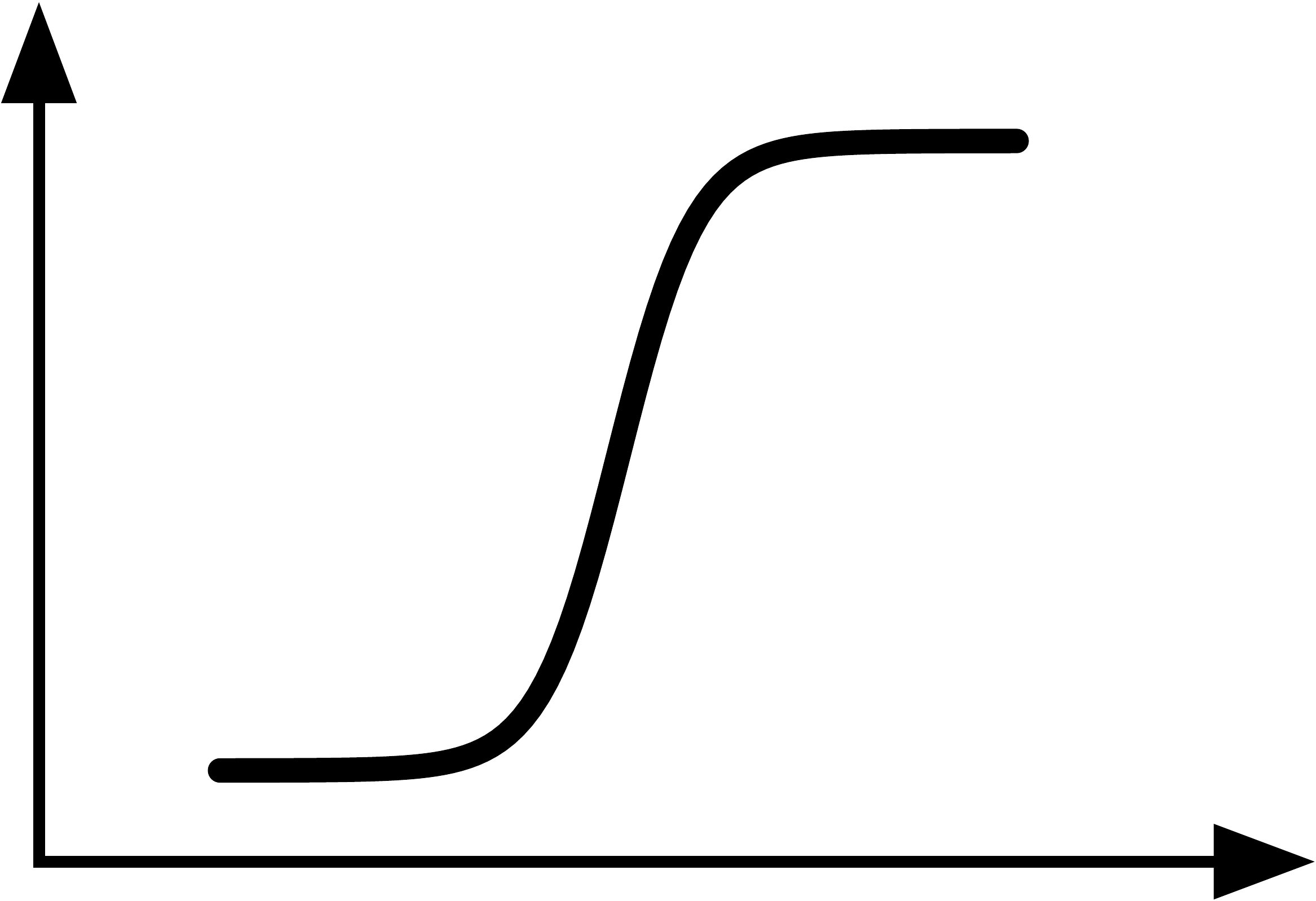}
\end{minipage}%
}\hspace{1pt}
\subfigure[Sharp SR]{
\begin{minipage}[b]{0.3\linewidth}
\includegraphics[width=1 \linewidth]{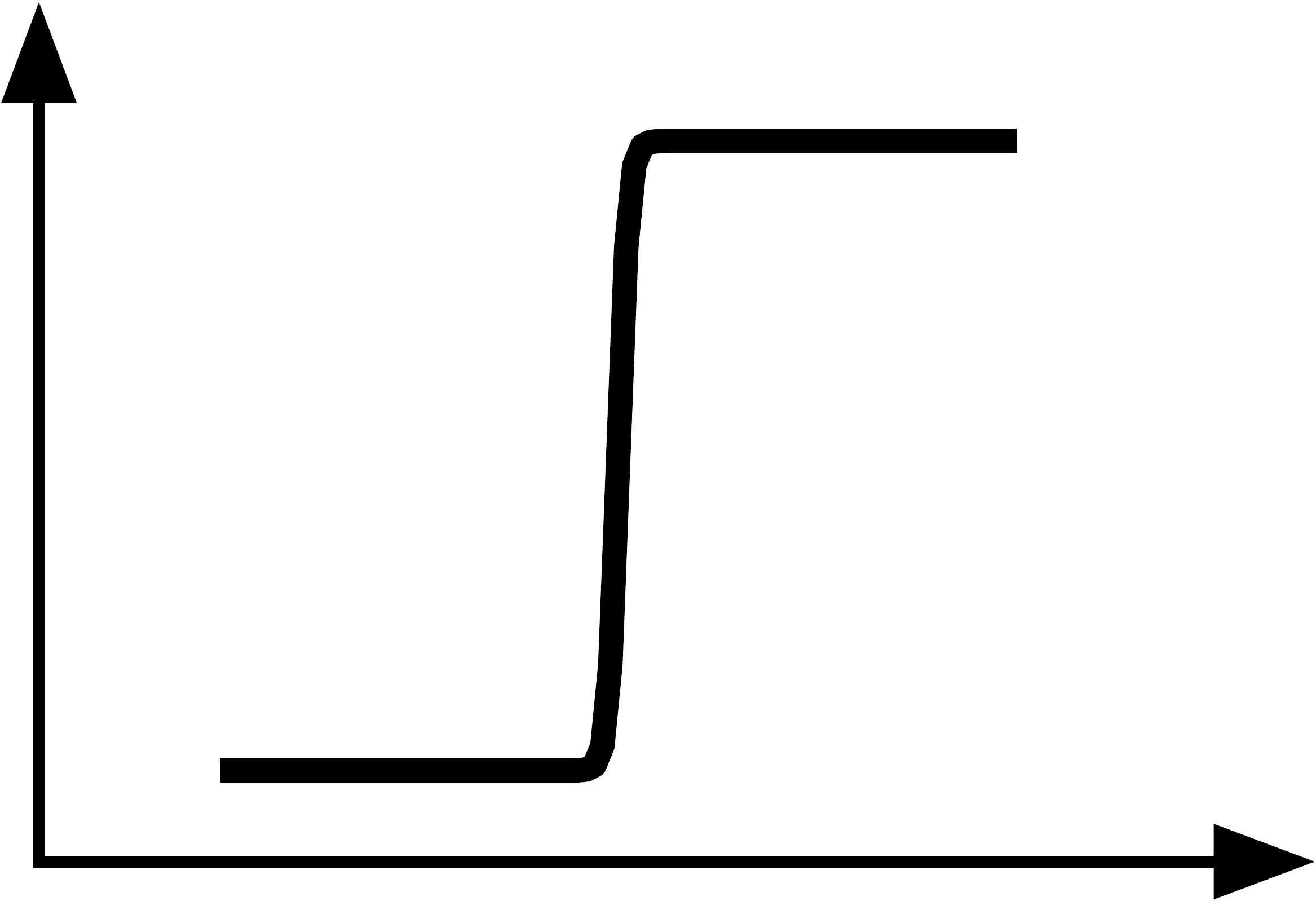}
\end{minipage}%
}\vspace{-2mm}

\subfigure[HR Gradiant]{
\begin{minipage}[b]{0.3\linewidth}
\includegraphics[width=1 \linewidth]{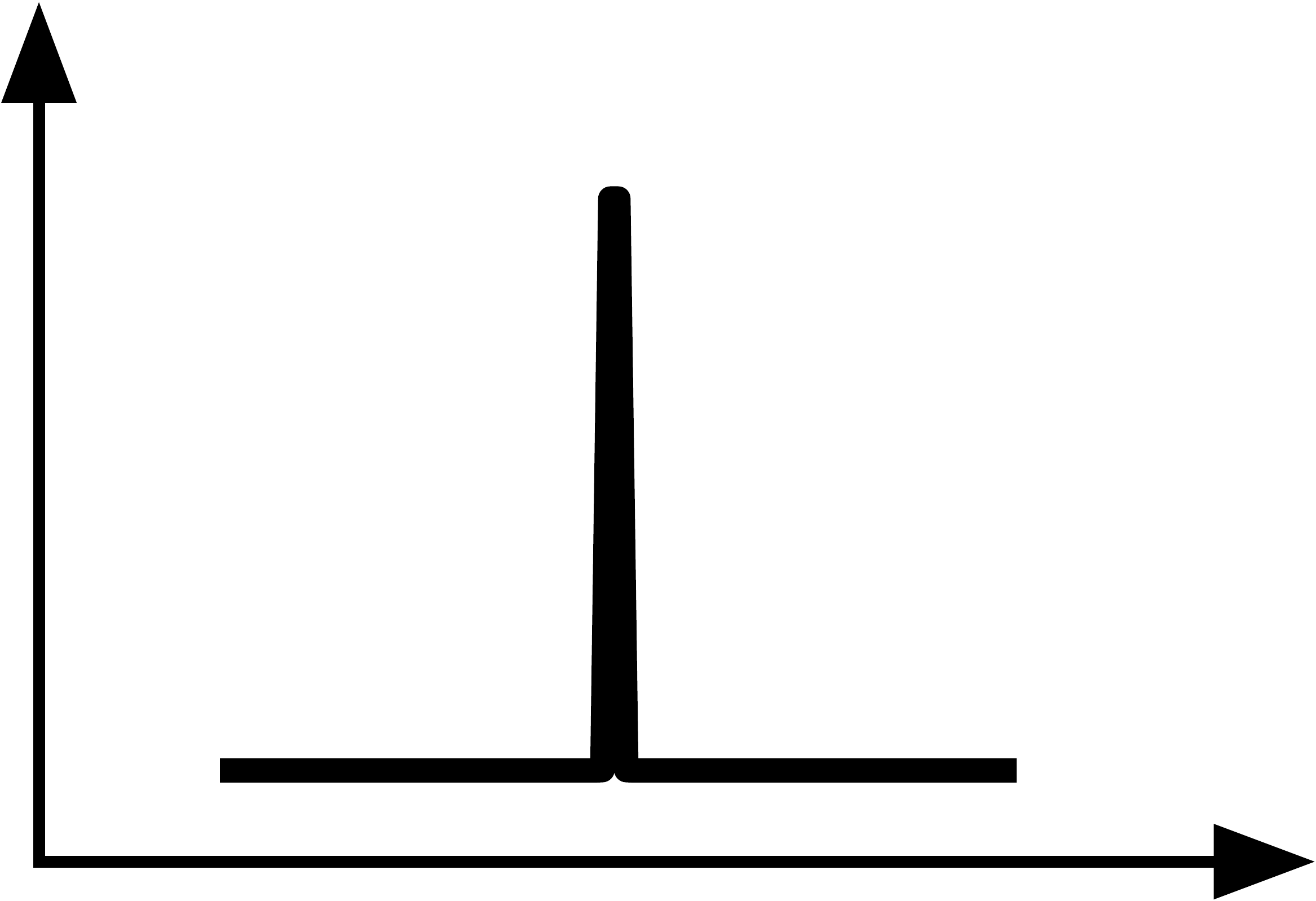}
\end{minipage}%
}\hspace{1pt}
\subfigure[Blurry Gradiant]{
\begin{minipage}[b]{0.3\linewidth}
\includegraphics[width=1 \linewidth]{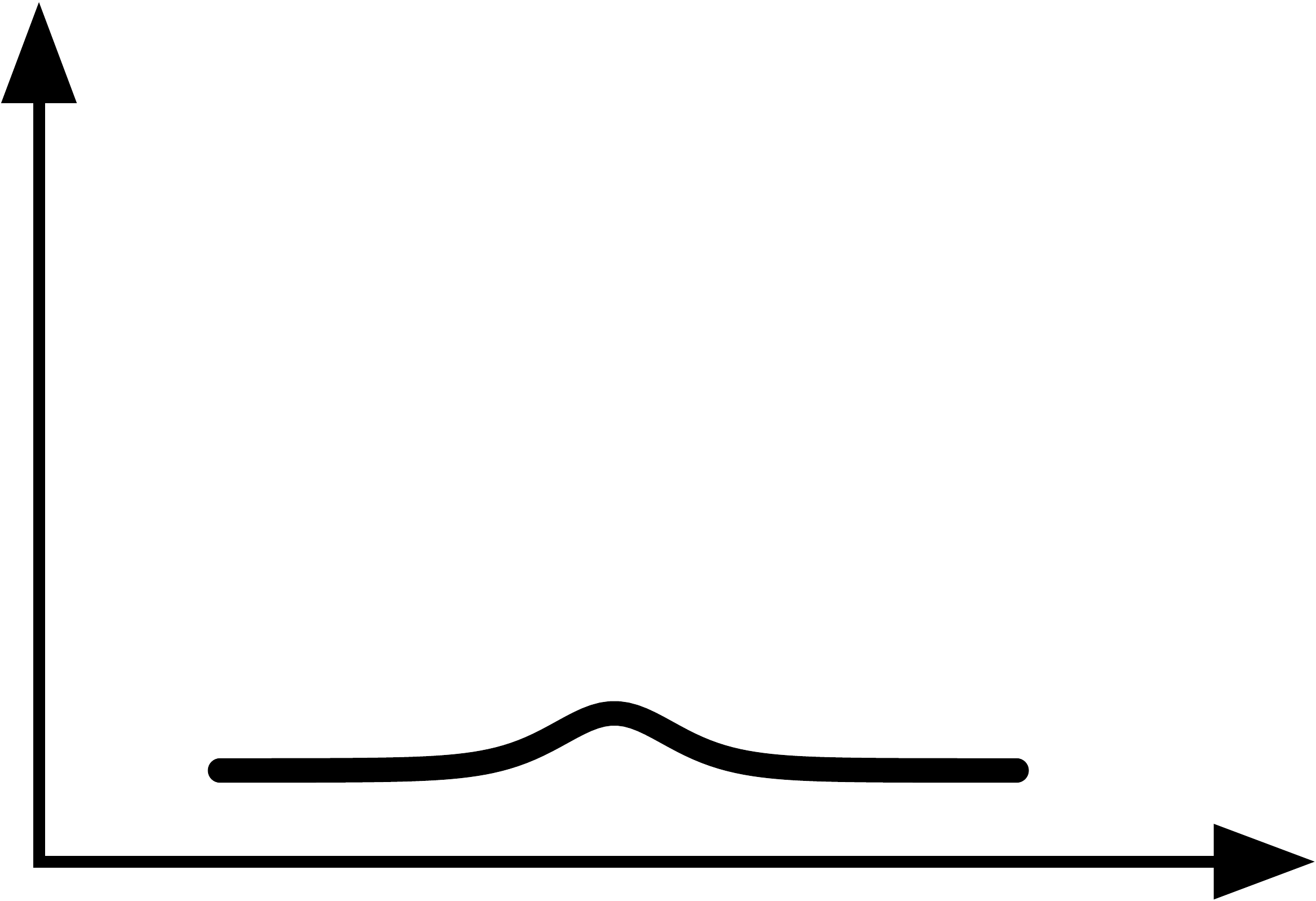}
\end{minipage}%
}\hspace{1pt}
\subfigure[Sharp Gradiant]{
\begin{minipage}[b]{0.3\linewidth}
\includegraphics[width=1 \linewidth]{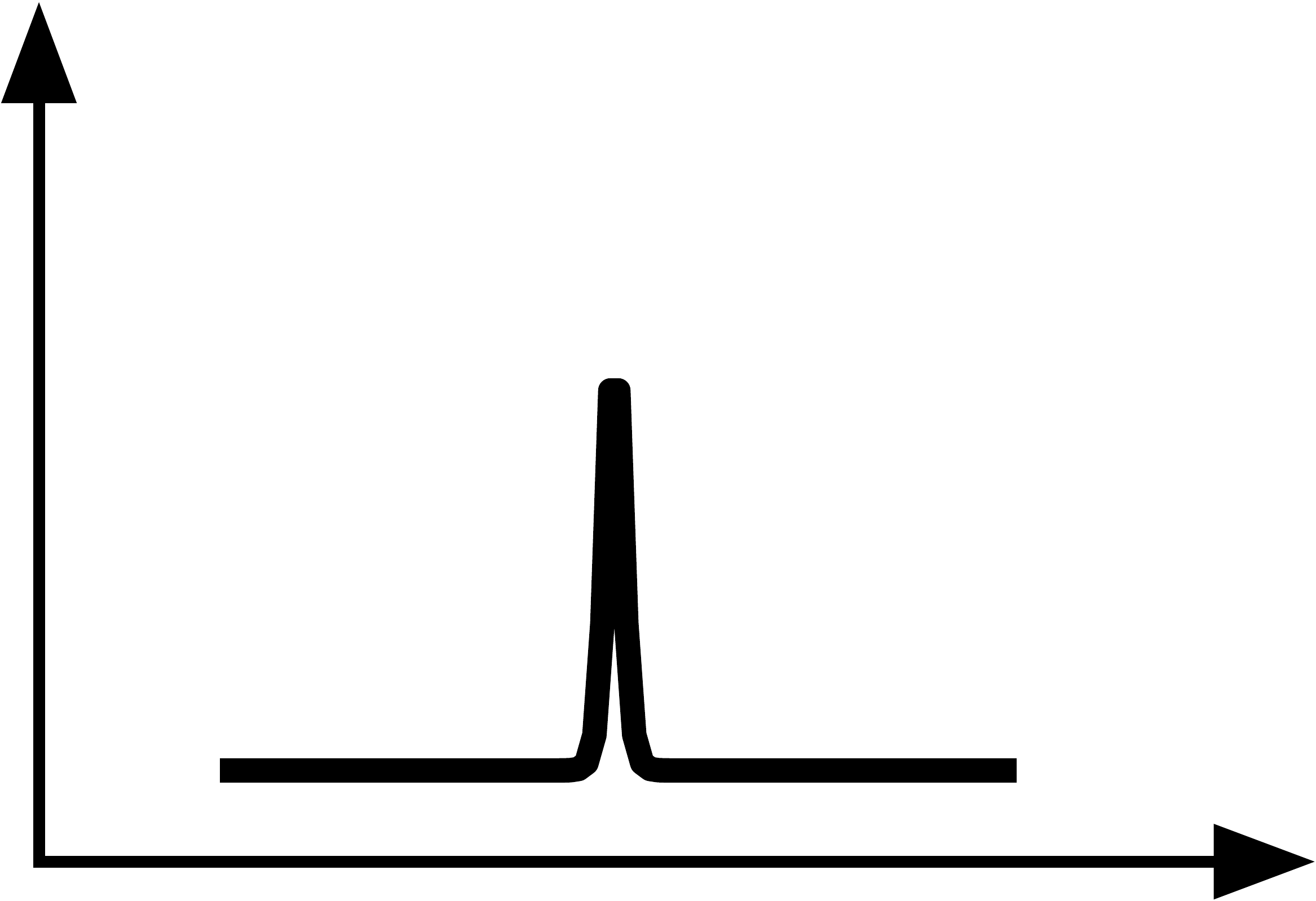}
\end{minipage}%
}

\centering
\caption{An illumination of a simple 1-D case. The first row shows the pixel sequences and the second row shows their corresponding gradient maps. }
\label{fig:method_graph}
\vspace{-2mm}
\end{figure}

\textbf{Gradient Loss}: Our motivation can be illustrated clearly by Figure~\ref{fig:method_graph}. Here we only consider a simple 1-dimensional case. If the model is only optimized in image space by the L1 loss, we usually get a SR sequence as Figure~\ref{fig:method_graph} (b) given an input testing sequence whose ground-truth is a sharp edge as Figure~\ref{fig:method_graph} (a). The model fails to recover sharp edges for the reason that the model tends to give an statistical average of possible HR solutions from training data. 
In this case, if we compute and show the gradient magnitudes of two sequences, it can be observed that the SR gradient is flat with low values while the HR gradient is a spike with high values. They are far from each other. This inspires us that if we add a second-order gradient constraint to the optimization objective, the model may learn more from the gradient space. It helps the model
focus on neighboring configuration, so that the local intensity of sharpness can be inferred more appropriately. Therefore, if the gradient information as Figure~\ref{fig:method_graph} (f) is captured, the probability of recovering Figure~\ref{fig:method_graph} (c) is increased significantly. SR methods can benefit from such guidance to avoid over-smooth or over-sharpening restoration. Moreover, it is easier to extract geometric characteristics in the gradient space. Hence geometric structures can be also preserved well, resulting in more photo-realistic SR images.

Here we propose a gradient loss to achieve the above goals. Since we have mentioned the gradient map is an ideal tool to reflect structural information of an image, it can also be utilized as a second-order constraint to provide supervision to the generator.  We formulate the gradient loss by diminishing the distance between the gradient map extracted from the SR image and the one from the corresponding HR image. 
With the supervision in both image and gradient domains, the generator can not only learn fine appearance, but also attach importance to avoiding detailed geometric distortions. Therefore, we design two terms of loss to penalize the difference in the gradient maps (GM) of the SR and HR images. One is based on the pixelwise loss as follows:
\begin{eqnarray}
\mathcal{L}^{{Pix}_{GM}}_{SR} = \mathbb{E}_{I^{SR}} \|M(G(I^{LR}))-M(I^{HR})\|_1. 
\end{eqnarray}
The other is to discriminate whether a gradient patch is from the HR gradient map. We design another gradient discriminator network to achieve this goal: 
\begin{eqnarray}
\mathcal{L}^{{Dis}_{GM}}_{SR} &=&  -\mathbb{E}_{I^{SR}}[\log (1-D_{GM}(M(I^{SR})))] \nonumber \\
&& -\mathbb{E}_{I^{HR}}[\log D_{GM}(M(I^{HR}))].
\end{eqnarray}
The gradient discriminator can also supervise the generation of SR results by adversarial learning:
\begin{eqnarray}
\mathcal{L}^{{Adv}_{GM}}_{SR} &=& -\mathbb{E}_{I^{SR}}[\log D_{GM}(M(G(I^{LR})))].
\end{eqnarray}

\begin{table*}
\centering
\caption{Comparison with state-of-the-art perceptual-driven SR methods on benchmark datasets. The best performance is \textbf{highlighted} in \textcolor{red}{\textbf{red}} (1st best) and \textcolor{blue}{\textbf{blue}} (2nd best). Our SPSR obtains the best PI and LPIPS values and comparable PSNR and SSIM values simultaneously. NatSR is more like a PSNR-oriented method since it has high PSNR and SSIM and relatively poor PI and LPIPS performance.  } \label{quantitative:PSNRPI}
\vspace{5px}
  \begin{tabular}{C{1.8cm}L{1.2cm}|C{1.8cm}C{1.9cm}C{1.9cm}C{1.9cm}C{1.9cm}|C{1.6cm}}
      \Xhline{1.0pt}
      \textbf{Dataset} & \textbf{Metric} & \textbf{Bicubic} & \textbf{SFTGAN}~\cite{SFTGAN} & \textbf{SRGAN}~\cite{ledig2017photo} & \textbf{ESRGAN}~\cite{wang2018esrgan} & \textbf{NatSR}~\cite{soh2019natural} & \textbf{SPSR} \\
      \Xhline{1.0pt}
      \multirow{4}*{\textbf{Set5}}
      & PI & 7.3699  & 3.7587  & 3.9820  & \textcolor{blue}{\textbf{3.7522}}  & 4.1648  & \textcolor{red}{\textbf{3.2743}}  \\ 
      & LPIPS & 0.3407  & 0.0890  & 0.0882  & \textcolor{blue}{\textbf{0.0748}}  & 0.0939  & \textcolor{red}{\textbf{0.0644}}  \\
      & PSNR & 28.420  & 29.932  & 29.168  & \textcolor{blue}{\textbf{30.454}}  & \textcolor{red}{\textbf{30.991}}  & 30.400  \\ 
      & SSIM & 0.8245  & 0.8665  & 0.8613  & \textcolor{blue}{\textbf{0.8677}}  & \textcolor{red}{\textbf{0.8800}}  & 0.8627  \\  \hline
      \multirow{4}*{\textbf{Set14}}
      & PI & 7.0268  & \textcolor{blue}{\textbf{2.9063}}  & 3.0851  & 2.9261  & 3.1094  & \textcolor{red}{\textbf{2.9036}}  \\ 
      & LPIPS & 0.4393  & 0.1481  & 0.1663  & \textcolor{blue}{\textbf{0.1329}}  & 0.1758  & \textcolor{red}{\textbf{0.1318}}  \\
      & PSNR & 26.100  & 26.223  & 26.171  & 26.276  & \textcolor{red}{\textbf{27.514}}  & \textcolor{blue}{\textbf{26.640}}  \\ 
      & SSIM & 0.7850  & 0.7854  & 0.7841  & 0.7783  & \textcolor{red}{\textbf{0.8140}}  & \textcolor{blue}{\textbf{0.7930}}  \\ \hline
      \multirow{4}*{\textbf{BSD100}}
      & PI & 7.0026  & \textcolor{blue}{\textbf{2.3774}}  & 2.5459  & 2.4793  & 2.7801  & \textcolor{red}{\textbf{2.3510}}  \\ 
      & LPIPS & 0.5249  & 0.1769  & 0.1980  & \textcolor{blue}{\textbf{0.1614}}  & 0.2114  & \textcolor{red}{\textbf{0.1611}}  \\
      & PSNR & 25.961  & 25.505  & 25.459  & 25.317  & \textcolor{red}{\textbf{26.445}}  & \textcolor{blue}{\textbf{25.505}}  \\ 
      & SSIM & 0.6675  & 0.6549  & 0.6485  & 0.6506  & \textcolor{red}{\textbf{0.6831}}  & \textcolor{blue}{\textbf{0.6576}}  \\ \hline
      \multirow{4}*{\textbf{General100}}
      & PI & 7.9365  & \textcolor{blue}{\textbf{4.2878}}  & 4.3757  & 4.3234  & 4.6262  & \textcolor{red}{\textbf{4.0991}}  \\ 
      & LPIPS & 0.3528  & 0.1030  & 0.1055  & \textcolor{blue}{\textbf{0.0879}}  & 0.1117  & \textcolor{red}{\textbf{0.0863}}  \\
      & PSNR & 28.018  & 29.026  & 28.575  & 29.412  & \textcolor{red}{\textbf{30.346}}  & \textcolor{blue}{\textbf{29.414}}  \\ 
      & SSIM & 0.8282  & 0.8508  & 0.8541  & \textcolor{blue}{\textbf{0.8546}}  & \textcolor{red}{\textbf{0.8721}}  & 0.8537  \\ \hline
      \multirow{4}*{\textbf{Urban100}}
      & PI & 6.9435  & \textcolor{blue}{\textbf{3.6136}}  & 3.6980  & 3.7704  & 3.6523  & \textcolor{red}{\textbf{3.5511}}  \\ 
      & LPIPS & 0.4726  & 0.1433  & 0.1551  & \textcolor{blue}{\textbf{0.1229}}  & 0.1500  & \textcolor{red}{\textbf{0.1184}}  \\
      & PSNR & 23.145  & 24.013  & 24.397  & 24.360  & \textcolor{red}{\textbf{25.464}}  & \textcolor{blue}{\textbf{24.799}}  \\ 
      & SSIM & 0.9011  & 0.9364  & 0.9381  & 0.9453  & \textcolor{red}{\textbf{0.9505}}  & \textcolor{blue}{\textbf{0.9481}} \\ \hline
      \Xhline{1.0pt}
  \end{tabular}
\vspace{-3mm}
\end{table*}

Note that each step in the operation $M(\cdot)$ is differentiable. Hence the model with gradient loss can be trained in an end-to-end manner. Furthermore, it is convenient to adopt gradient loss as additional guidance in any generative model due to the concise formulation and strong transferability.  

\textbf{Overall Objective}: In conclusion, we have two discriminators $D_I$ and $D_{GM}$ which are optimized by $\mathcal{L}^{{Dis}_I}_{SR}$ and $\mathcal{L}^{{Dis}_{GM}}_{SR}$, respectively. For the generator, two terms of loss are used to provide supervision signals simultaneously. One is imposed on the structure-preserving SR branch while the other is to reconstruct high-quality gradient maps by minimizing the pixelwise loss $\mathcal{L}^{Pix_{GM}}_{GB}$ in the gradient branch (GB).  
The overall objective is defined as follows: 
\begin{eqnarray}
\mathcal{L}^{G} &=& \mathcal{L}^{G}_{SR}+\mathcal{L}^{G}_{GB} \notag \\ 
&=&\mathcal{L}^{Per}_{SR}+\beta^I_{SR}\mathcal{L}^{Pix_I}_{SR}+\gamma^I_{SR}\mathcal{L}^{{Adv}_I}_{SR}+\beta^{GM}_{SR}\mathcal{L}^{Pix_{GM}}_{SR} \notag \\ 
&&+\gamma^{GM}_{SR}\mathcal{L}^{Adv_{GM}}_{SR}+\beta^{GM}_{GB}\mathcal{L}^{Pix_{GM}}_{GB}. 
\end{eqnarray}
$\beta^I_{SR}$, $\gamma^I_{SR}$, $\beta^{GM}_{SR}$, $\gamma^{GM}_{SR}$ and $\beta^{GM}_{GB}$ denote the trade-off parameters of different losses. Among these, $\beta^I_{SR}$, $\beta^{GM}_{SR}$ and $\beta^{GM}_{GB}$ are the weights of the pixel losses for SR images, gradient maps of SR images and SR gradient maps respectively. $\gamma^I_{SR}$ and $\gamma^{GM}_{SR}$ are the weights of the adversarial losses for SR image and their gradient maps.

\section{Experiments}

\subsection{Implementation Details}

\textbf{Datasets and Evaluation Metrics}: We evaluate the SR performance of our proposed SPSR method. We utilize DIV2K~\cite{agustsson2017ntire} as the training dataset and five commonly used benchmarks for testing: Set5~\cite{bevilacqua2012low}, Set14~\cite{zeyde2010single}, BSD100~\cite{BSD100}, Urban100~\cite{huang2015single} and General100~\cite{dong2016accelerating}. We downsample HR images by bicubic interpolation to get LR inputs and only consider the scaling factor of $4\times$ in our experiments. We choose Perceptual Index (PI)~\cite{blau20182018}, Learned Perceptual Image Patch Similarity (LPIPS)~\cite{zhang2018unreasonable}, PSNR and Structure Similarity (SSIM)~\cite{wang2004image} as the evaluation metrics. Lower PI and LPIPS values indicate higher perceptual quality.

\textbf{Training Details}: We use the architecture of ESRGAN~\cite{wang2018esrgan} as the backbone of our SR branch and the RRDB block~\cite{wang2018esrgan} as the gradient block. We randomly sample 15 $32\times32$ patches from LR images for each input mini-batch. Therefore the ground-truth HR patches have a size of $128\times128$. We initialize the generator with the parameters of a pre-trained PSNR-oriented model. The pixelwise loss, perceptual loss, adversarial loss and gradient loss are used as the optimizing objectives. A pre-trained 19-layer VGG network~\cite{simonyan2014very} is employed to calculate the feature distances in the perceptual loss. We also use a VGG-style network to perform discrimination. 
ADAM optimizor~\cite{Adam} with $\beta_1=0.9, \beta_2=0.999$ and $\epsilon=1\times10^{-8}$ is used for optimization. We set the learning rates to $1\times10^{-4}$ for both generator and discriminator, and reduce them to half at 50k, 100k, 200k, 300k iterations. As for the trade-off parameters of losses, we follow the settings in~\cite{wang2018esrgan} and set $\beta^I_{SR}$ and $\gamma^I_{SR}$ to $0.01$ and $0.005$, accordingly. Then we set the weights of gradient loss equal to those of image-space loss. Hence $\beta^{GM}_{SR}=0.01$ and $\gamma^{GM}_{SR}=0.005$. In terms of $\beta^{GM}_{GB}$, we set it to $0.5$ for better performance of gradient translation. 
All the experiments are implemented by PyTorch~\cite{PyTorch} on NVIDIA GTX 1080Ti GPUs. 

\subsection{Results and Analysis}

\textbf{Quantitative Comparison}: We compare our method quantitatively with state-of-the-art perceptual-driven SR methods including SFTGAN~\cite{SFTGAN}, SRGAN~\cite{ledig2017photo}, ESRGAN~\cite{wang2018esrgan} and NatSR~\cite{soh2019natural}. Results of PI, LPIPS, PSNR and SSIM values are presented in Table~\ref{quantitative:PSNRPI}. In each row, the best result is highlighted in red while the second best is in blue. We can see in all the testing datasets SPSR achieves the best PI and LPIPS performance. Meanwhile, we get the second best PSNR and SSIM values in most datasets. It is noteworthy that while NatSR gets the highest PSNR and SSIM values in all the datasets, our method surpasses NatSR by a large margin in terms of PI and LPIPS. Moreover, NatSR cannot achieve the second best PI and LPIPS values in any testing set. Thus NatSR is more like a PSNR-oriented SR method, which tends to produce relatively blurry results with high PSNR compared to other perceptual-driven methods. Besides, we get better performance than ESRGAN with only a little increment on network parameters in the SR branch. Therefore, the results demonstrate the superior ability of our SPSR method to obtain excellent perceptual quality and minor distortions simultaneously.

\begin{figure*}[htbp]
   \newlength\fsdttwofig
   \setlength{\fsdttwofig}{-1.5mm}
   \scriptsize
   \centering
   
   \begin{tabular}{cc}
      
      \begin{adjustbox}{valign=t}
      \tiny
         \begin{tabular}{c}
            \includegraphics[width=0.25\textwidth]{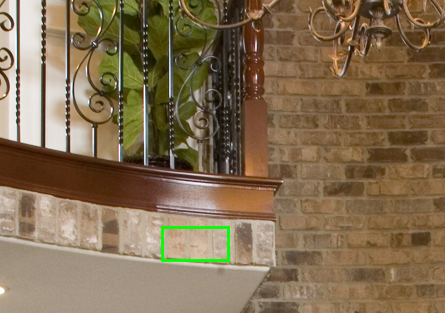}
            \\
             `im\_004' from General100
            
         \end{tabular}
      \end{adjustbox}
      \hspace{-2.3mm}
      \begin{adjustbox}{valign=t}
      \tiny
         \begin{tabular}{ccccc}
            \includegraphics[width=\widthscalefive \textwidth]{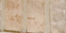} \hspace{\fsdttwofig} &
            \includegraphics[width=\widthscalefive \textwidth]{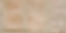} \hspace{\fsdttwofig} &
            \includegraphics[width=\widthscalefive \textwidth]{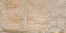} \hspace{\fsdttwofig} &
            \includegraphics[width=\widthscalefive \textwidth]{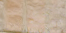} 
            \\
            HR \hspace{\fsdttwofig} &
            Bicubic \hspace{\fsdttwofig} &
            EnhanceNet \hspace{\fsdttwofig} &
            SFTGAN
            \\
            \includegraphics[width=\widthscalefive \textwidth]{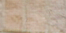} \hspace{\fsdttwofig} &
            \includegraphics[width=\widthscalefive \textwidth]{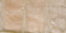} \hspace{\fsdttwofig} &
            \includegraphics[width=\widthscalefive \textwidth]{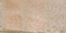} \hspace{\fsdttwofig} &
            \includegraphics[width=\widthscalefive \textwidth]{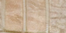} 
            \\ 
            SRGAN \hspace{\fsdttwofig} &
            ESRGAN \hspace{\fsdttwofig} &
            NatSR \hspace{\fsdttwofig} &
            SPSR
            \\
         \end{tabular}
      \end{adjustbox}
      \vspace{0.5mm}
      \\
      
      \begin{adjustbox}{valign=t}
      \tiny
         \begin{tabular}{c}
            \includegraphics[width=0.25\textwidth]{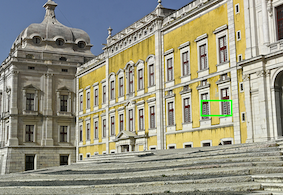}
            \\
             `img\_054' from Urban100
            
         \end{tabular}
      \end{adjustbox}
      \hspace{-2.3mm}
      \begin{adjustbox}{valign=t}
      \tiny
         \begin{tabular}{ccccc}
            \includegraphics[width=\widthscalefive \textwidth]{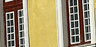} \hspace{\fsdttwofig} &
            \includegraphics[width=\widthscalefive \textwidth]{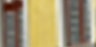} \hspace{\fsdttwofig} &
            \includegraphics[width=\widthscalefive \textwidth]{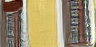} \hspace{\fsdttwofig} &
            \includegraphics[width=\widthscalefive \textwidth]{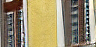} 
            \\
            HR \hspace{\fsdttwofig} &
            Bicubic \hspace{\fsdttwofig} &
            EnhanceNet \hspace{\fsdttwofig} &
            SFTGAN
            \\
            \includegraphics[width=\widthscalefive \textwidth]{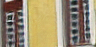} \hspace{\fsdttwofig} &
            \includegraphics[width=\widthscalefive \textwidth]{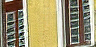} \hspace{\fsdttwofig} &
            \includegraphics[width=\widthscalefive \textwidth]{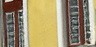} \hspace{\fsdttwofig} &
            \includegraphics[width=\widthscalefive \textwidth]{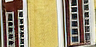} 
            \\ 
            SRGAN \hspace{\fsdttwofig} &
            ESRGAN \hspace{\fsdttwofig} &
            NatSR \hspace{\fsdttwofig} &
            SPSR
            \\
         \end{tabular}
      \end{adjustbox}
      \vspace{0.5mm}
      \\

      \begin{adjustbox}{valign=t}
      \tiny
         \begin{tabular}{c}
            \includegraphics[width=0.25\textwidth]{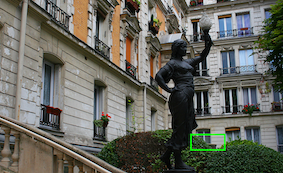}
            \\
            `img\_003' from Urban100
         \end{tabular}
      \end{adjustbox}
      \hspace{-2.3mm}
      \begin{adjustbox}{valign=t}
      \tiny
         \begin{tabular}{ccccc}
            \includegraphics[width=\widthscalefive \textwidth]{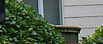} \hspace{\fsdttwofig} &
            \includegraphics[width=\widthscalefive \textwidth]{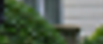} \hspace{\fsdttwofig} &
            \includegraphics[width=\widthscalefive \textwidth]{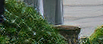} \hspace{\fsdttwofig} &
            \includegraphics[width=\widthscalefive \textwidth]{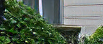} 
            \\
            HR \hspace{\fsdttwofig} &
            Bicubic \hspace{\fsdttwofig} &
            EnhanceNet \hspace{\fsdttwofig} &
            SFTGAN
            \\
            \includegraphics[width=\widthscalefive \textwidth]{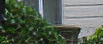} \hspace{\fsdttwofig} &
            \includegraphics[width=\widthscalefive \textwidth]{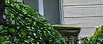} \hspace{\fsdttwofig} &
            \includegraphics[width=\widthscalefive \textwidth]{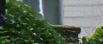} \hspace{\fsdttwofig} &
            \includegraphics[width=\widthscalefive \textwidth]{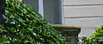} 
            \\ 
            SRGAN \hspace{\fsdttwofig} &
            ESRGAN \hspace{\fsdttwofig} &
            NatSR \hspace{\fsdttwofig} &
            SPSR
            \\
         \end{tabular}
      \end{adjustbox}
      \vspace{0.5mm}
      \\
      
      \begin{adjustbox}{valign=t}
      \tiny
         \begin{tabular}{c}
            \includegraphics[width=0.25\textwidth]{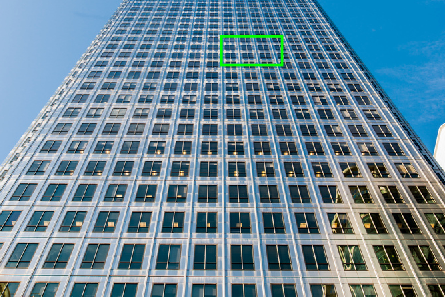}
            \\
            `img\_030' from Urban100
         \end{tabular}
      \end{adjustbox}
      \hspace{-2.3mm}
      \begin{adjustbox}{valign=t}
      \tiny
         \begin{tabular}{ccccc}
            \includegraphics[width=\widthscalefive \textwidth]{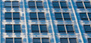} \hspace{\fsdttwofig} &
            \includegraphics[width=\widthscalefive \textwidth]{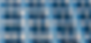} \hspace{\fsdttwofig} &
            \includegraphics[width=\widthscalefive \textwidth]{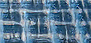} \hspace{\fsdttwofig} &
            \includegraphics[width=\widthscalefive \textwidth]{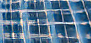} 
            \\
            HR \hspace{\fsdttwofig} &
            Bicubic \hspace{\fsdttwofig} &
            EnhanceNet \hspace{\fsdttwofig} &
            SFTGAN
            \\
            \includegraphics[width=\widthscalefive \textwidth]{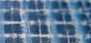} \hspace{\fsdttwofig} &
            \includegraphics[width=\widthscalefive \textwidth]{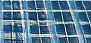} \hspace{\fsdttwofig} &
            \includegraphics[width=\widthscalefive \textwidth]{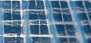} \hspace{\fsdttwofig} &
            \includegraphics[width=\widthscalefive \textwidth]{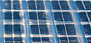} 
            \\ 
            SRGAN \hspace{\fsdttwofig} &
            ESRGAN \hspace{\fsdttwofig} &
            NatSR \hspace{\fsdttwofig} &
            SPSR
            \\
         \end{tabular}
      \end{adjustbox}
      
   \end{tabular}\vspace{1mm}
   \caption{
      Visual comparison with state-of-the-art perceptual-driven SR methods. The results show that our proposed SPSR method significantly outperforms other methods in structure restoration while generating perceptual-pleasant SR images. Best viewed on screen.  
   }
\label{fig:best_result_4x}
\vspace{-2mm}
\end{figure*}

\begin{figure*}[htbp]
   \setlength{\fsdttwofig}{-1.5mm}
   \scriptsize
   \centering
   
   \begin{tabular}{cc}
      
      \begin{adjustbox}{valign=t}
      \tiny
         \begin{tabular}{c}
            \includegraphics[width=0.25\textwidth]{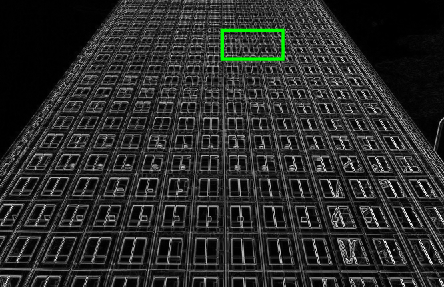}
            \\
             `im\_030' from Urban100
            
         \end{tabular}
      \end{adjustbox}
      \hspace{-2.3mm}
      \begin{adjustbox}{valign=t}
      \tiny
         \begin{tabular}{ccccc}
            \includegraphics[width=\widthscalefive \textwidth]{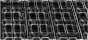} \hspace{\fsdttwofig} &
            \includegraphics[width=\widthscalefive \textwidth]{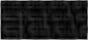} \hspace{\fsdttwofig} &
            \includegraphics[width=\widthscalefive \textwidth]{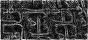} \hspace{\fsdttwofig} &
            \includegraphics[width=\widthscalefive \textwidth]{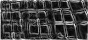} 
            \\
            HR \hspace{\fsdttwofig} &
            Bicubic \hspace{\fsdttwofig} &
            EnhanceNet \hspace{\fsdttwofig} &
            SFTGAN
            \\
            \includegraphics[width=\widthscalefive \textwidth]{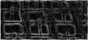} \hspace{\fsdttwofig} &
            \includegraphics[width=\widthscalefive \textwidth]{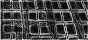} \hspace{\fsdttwofig} &
            \includegraphics[width=\widthscalefive \textwidth]{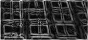} \hspace{\fsdttwofig} &
            \includegraphics[width=\widthscalefive \textwidth]{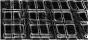} 
            \\ 
            SRGAN \hspace{\fsdttwofig} &
            ESRGAN \hspace{\fsdttwofig} &
            NatSR \hspace{\fsdttwofig} &
            SPSR
            \\
         \end{tabular}
      \end{adjustbox}
      
   \end{tabular}\vspace{1mm}
   \caption{
      Comparison of gradient maps with state-of-the-art perceptual-driven SR methods. The proposed SPSR method can better preserve gradients and structures. Best viewed on screen. 
   }
\label{fig:best_result_4x_grad}
\vspace{-2mm}
\end{figure*}

\textbf{Qualitative Comparison}: We also conduct visual comparison to perceptual-driven SR methods. From Figure~\ref{fig:best_result_4x} we see that our results are more natural and realistic than other methods. For the first image, SPSR infers sharp edges of the bricks properly, indicating that our method is capable of capturing structural characteristics of objects in images. In other rows, our method also recovers better textures than the compared SR methods. The structures in our results are clear without severe distortions, while other methods fail to show satisfactory appearance for the objects. Gradient maps for the last row are shown in Figure~\ref{fig:best_result_4x_grad}. We can see the gradient maps of other methods tend to have small values or contain structure degradation while ours are bold and natural. 
The qualitative comparison proves that our proposed SPSR method can learn more structure information from the gradient space, which helps generate photo-realistic SR images by preserving geometric structures.

\begin{figure}[t]
\centering

\subfigure[HR]{
\begin{minipage}[b]{0.48\linewidth}
\includegraphics[width=1 \linewidth]{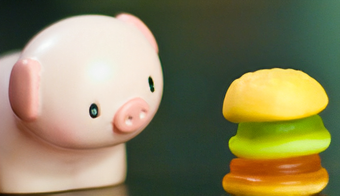}
\end{minipage}%
}%
\subfigure[HR gradiant]{
\begin{minipage}[b]{0.48\linewidth}
\includegraphics[width=1 \linewidth]{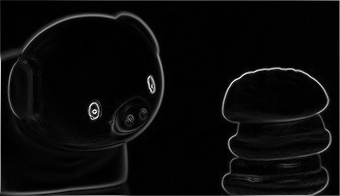}
\end{minipage}%
}%
\vspace{-3mm}

\subfigure[LR gradiant (Bicubic)]{
\begin{minipage}[b]{0.48\linewidth}
\includegraphics[width=1 \linewidth]{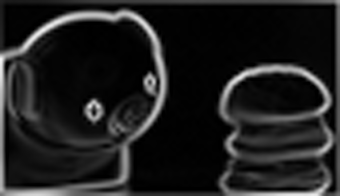}
\end{minipage}%
}%
\subfigure[Output of the gradiant branch]{
\begin{minipage}[b]{0.48\linewidth}
\includegraphics[width=1 \linewidth]{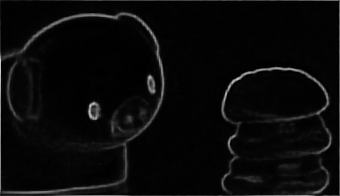}
\end{minipage}%
}%

\centering
\caption{Visualization of gradient maps (`im\_073' from General100). The HR gradient map has thin outlines while those in the LR gradient map are thick. Our gradient branch is able to recover HR gradient maps with pleasant structures. }
\label{fig:branch_result_4x}
\vspace{-5mm}
\end{figure}

\begin{figure}[t]
\centering

\subfigure[Only the SR branch]{
\begin{minipage}[b]{0.48\linewidth}
\includegraphics[width=1 \linewidth]{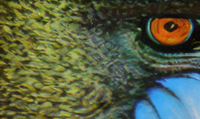}\vspace{1.8pt}
\includegraphics[width=1 \linewidth]{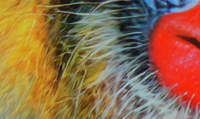}
\end{minipage}%
}%
\subfigure[Complete model]{
\begin{minipage}[b]{0.48\linewidth}
\includegraphics[width=1 \linewidth]{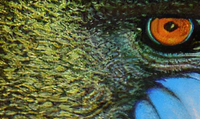}\vspace{1.8pt}
\includegraphics[width=1 \linewidth]{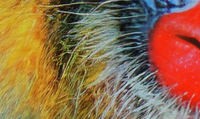}
\end{minipage}
}

\centering
\caption{SR comparison of the models without and with the gradient branch (`baboon' from Set14). Images recovered by the complete model have clearer textures than those generated only by the features from the SR branch. }
\label{fig:branch_region_result_4x}
\vspace{-3mm}
\end{figure}

\begin{table*}
\centering
\caption{Comparison of models with different components. The best results are \textbf{highlighted}. SPSR w/o GB has better PI performance than ESRGAN in all the benchmark datasets. SPSR surpasses ESRGAN on all the measurements in all the testing sets. } \label{table:Comparison}
 \vspace{5px}
  \begin{tabular}{C{2.5cm}|C{1.1cm}C{1.1cm}C{1.1cm}|C{1.1cm}C{1.1cm}C{1.1cm}|C{1.1cm}C{1.1cm}C{1.1cm}}
  \Xhline{1.0pt}
  \multirow{2}*{\textbf{Method}} & 
  \multicolumn{3}{c|}{\textbf{Set14}} & 
  \multicolumn{3}{c|}{\textbf{BSD100}} & 
  \multicolumn{3}{c}{\textbf{Urban100}} \\
  &
  \textbf{PI} &
  \textbf{PSNR} &
  \textbf{SSIM} &
  \textbf{PI} &
  \textbf{PSNR} &
  \textbf{SSIM} &
  \textbf{PI} &
  \textbf{PSNR} &
  \textbf{SSIM} \\
      \Xhline{1.0pt}
      \multirow{1}*{\textbf{ESRGAN}~\cite{wang2018esrgan}} & 
      \multirow{1}*{2.926}  & 
      \multirow{1}*{26.276}  & 
      \multirow{1}*{0.778}  & 
      \multirow{1}*{2.479}  & 
      \multirow{1}*{25.317}  & 
      \multirow{1}*{0.651}  & 
      \multirow{1}*{3.770}   & 
      \multirow{1}*{24.360}  & 
      \multirow{1}*{0.945}  \\ 
      \hline
      \multirow{1}*{\textbf{SPSR w/o GB}} & 
      \multirow{1}*{\textbf{2.864}}  & 
      \multirow{1}*{26.027}  & 
      \multirow{1}*{0.785}  & 
      \multirow{1}*{2.370}  & 
      \multirow{1}*{25.376}  & 
      \multirow{1}*{\textbf{0.659}}  & 
      \multirow{1}*{3.604}  &
      \multirow{1}*{23.939}  & 
      \multirow{1}*{0.940}  \\ 
      \hline
      \multirow{1}*{\textbf{SPSR w/o GL}} & 
      \multirow{1}*{3.028}  & 
      \multirow{1}*{26.547}  & 
      \multirow{1}*{\textbf{0.794}}  & 
      \multirow{1}*{2.456}  & 
      \multirow{1}*{25.214}  & 
      \multirow{1}*{0.647}  & 
      \multirow{1}*{3.605}  & 
      \multirow{1}*{24.309}  & 
      \multirow{1}*{0.942}  \\ 
      \hline
      \multirow{1}*{\textbf{SPSR}} & 
      \multirow{1}*{2.904}  & 
      \multirow{1}*{\textbf{26.640}}  & 
      \multirow{1}*{0.793}  & 
      \multirow{1}*{\textbf{2.351}}  & 
      \multirow{1}*{\textbf{25.505}}  & 
      \multirow{1}*{0.658}  & 
      \multirow{1}*{\textbf{3.551}}  &
      \multirow{1}*{\textbf{24.799}}  & 
      \multirow{1}*{\textbf{0.948}} \\ 
      \hline
\Xhline{1.0pt}
  \end{tabular}
\vspace{-3mm}
\end{table*}

\textbf{User Study}: We further perform a user study to evaluate visual quality of different SR methods. Detailed settings and results are presented in the supplementary material. 

\textbf{Ablation Study}: We conduct more experiments on different models to validate the necessity of each part in our proposed framework. 
Since we apply the architecture of ESRGAN~\cite{wang2018esrgan} in our SR branch, we use ESRGAN as the baseline. We compare three models with it. The first one has the same architecture as ESRGAN without the gradient branch (GB) and is trained by both the image-space and gradient-space loss. 
The second one is trained without the gradient loss (GL), but has the gradient branch in the network. The third is our proposed SPSR model, utilizing both the gradient loss and the gradient branch. Quantitative comparison is presented in Table~\ref{table:Comparison}. It is observed that SPSR w/o GB has a significant enhancement on PI performance over ESRGAN, which demonstrates the effectiveness of the proposed gradient loss in improving perceptual quality. Besides, the results of SPSR w/o GL also show that the gradient branch can significantly help improve PI or PSNR while relatively preserving the other one. In terms of the complete model, we can see SPSR surpasses ESRGAN on all the measurements in all the testing sets. Therefore, the effectiveness of our method is verified clearly. 

\textbf{Effects of the Gradient Branch}: In order to validate the effectiveness of the gradient branch, we also visualize the output gradient maps as shown in Figure~\ref{fig:branch_result_4x}. Given HR images with sharp edges, the extracted HR gradient maps may have thin and clear outlines for objects in the images. However, the gradient maps extracted from the LR counterparts commonly have thick lines after the bicubic upsampling. Our gradient branch takes LR gradient maps as inputs and produce HR gradient maps so as to provide explicit structural information as a guidance for the SR branch. By treating gradient generation as an image translation problem, we can exploit the strong generative ability of the deep model. From the output gradient map in Figure~\ref{fig:branch_result_4x} (d), we can see our gradient branch successfully recover thin and structure-pleasing gradient maps. 

We conduct another experiment to evaluate the effectiveness of the gradient branch. With a complete SPSR model, we remove the features from the gradient branch by setting them to $0$ and only use the SR branch for inference. The visualization results are shown in Figure~\ref{fig:branch_region_result_4x}. From the patches, we can see the furs and whiskers super-resolved by only the SR branch are more blurry than those recovered by the complete model. The change of detailed textures reveals that the gradient branch can help produce sharp edges for better perceptual fidelity.

\section{Conclusion}

In this paper, we have proposed a structure-preserving super resolution method (SPSR) with gradient guidance to alleviate the issue of geometric distortions commonly existing in the SR results of perceptual-driven methods. We have preserved geometric structures in two aspects. Firstly, we build a gradient branch which aims to recover high-resolution gradient maps from the LR ones and provides gradient information to the SR branch as an explicit structural guidance. Secondly, we propose a new gradient loss to impose second-order restrictions on the recovered images. Geometric relationship can be better captured with both the image-space and gradient-space supervision. Quantitative and qualitative experimental results on five popular benchmark testing sets have shown the effectiveness of our proposed method.

\section*{Acknowledgement}
This work was supported in part by the National Key Research and Development Program of China under Grant 2017YFA0700802, in part by the National Natural Science Foundation of China under Grant 61822603, Grant U1813218, Grant U1713214, and Grant 61672306, in part by the Shenzhen Fundamental Research Fund (Subject Arrangement) under Grant JCYJ20170412170602564, and in part by Tsinghua University Initiative Scientfic Research Program.

{\small
\bibliographystyle{ieee_fullname}
\bibliography{egbib}

\begin{thebibliography}{10}\itemsep=-1pt

\bibitem{agustsson2017ntire}
Eirikur Agustsson and Radu Timofte.
\newblock Ntire 2017 challenge on single image super-resolution: Dataset and
  study.
\newblock In {\em CVPR}, pages 126--135, 2017.

\bibitem{anoosheh2019night}
Asha Anoosheh, Torsten Sattler, Radu Timofte, Marc Pollefeys, and Luc Van~Gool.
\newblock Night-to-day image translation for retrieval-based localization.
\newblock In {\em ICRA}, pages 5958--5964. IEEE, 2019.

\bibitem{arjovsky2017wasserstein}
Martin Arjovsky, Soumith Chintala, and L{\'e}on Bottou.
\newblock Wasserstein gan.
\newblock {\em arXiv preprint arXiv:1701.07875}, 2017.

\bibitem{berthelot2017began}
David Berthelot, Thomas Schumm, and Luke Metz.
\newblock Began: Boundary equilibrium generative adversarial networks.
\newblock {\em arXiv preprint arXiv:1703.10717}, 2017.

\bibitem{bevilacqua2012low}
Marco Bevilacqua, Aline Roumy, Christine Guillemot, and Marie-Line
  Alberi-Morel.
\newblock Low-complexity single-image super-resolution based on nonnegative
  neighbor embedding.
\newblock In {\em BMVC}, 2012.

\bibitem{blau20182018}
Yochai Blau, Roey Mechrez, Radu Timofte, Tomer Michaeli, and Lihi Zelnik-Manor.
\newblock The 2018 pirm challenge on perceptual image super-resolution.
\newblock In {\em ECCV}, pages 334--355. Springer, 2018.

\bibitem{classicalNeighborEmbedding}
Hong Chang, Dit{-}Yan Yeung, and Yimin Xiong.
\newblock Super-resolution through neighbor embedding.
\newblock In {\em CVPR}, pages 275--282, 2004.

\bibitem{dong2014learning}
Chao Dong, Chen~Change Loy, Kaiming He, and Xiaoou Tang.
\newblock Learning a deep convolutional network for image super-resolution.
\newblock In {\em ECCV}, pages 184--199. Springer, 2014.

\bibitem{dong2016accelerating}
Chao Dong, Chen~Change Loy, and Xiaoou Tang.
\newblock Accelerating the super-resolution convolutional neural network.
\newblock In {\em ECCV}, pages 391--407. Springer, 2016.

\bibitem{duchon1979lanczos}
Claude~E Duchon.
\newblock Lanczos filtering in one and two dimensions.
\newblock {\em Journal of applied meteorology}, 18(8):1016--1022, 1979.

\bibitem{fattal2007image}
Raanan Fattal.
\newblock Image upsampling via imposed edge statistics.
\newblock {\em TOG}, 26(3):95, 2007.

\bibitem{freedman2011image}
Gilad Freedman and Raanan Fattal.
\newblock Image and video upscaling from local self-examples.
\newblock {\em TOG}, 30(2):12, 2011.

\bibitem{classicalExampleBased}
William~T. Freeman, Thouis~R. Jones, and Egon~C. Pasztor.
\newblock Example-based super-resolution.
\newblock {\em CG\&A}, 22(2):56--65, 2002.

\bibitem{glasner2009super}
Daniel Glasner, Shai Bagon, and Michal Irani.
\newblock Super-resolution from a single image.
\newblock In {\em ICCV}, pages 349--356. IEEE, 2009.

\bibitem{goodfellow2014generative}
Ian Goodfellow, Jean Pouget-Abadie, Mehdi Mirza, Bing Xu, David Warde-Farley,
  Sherjil Ozair, Aaron Courville, and Yoshua Bengio.
\newblock Generative adversarial nets.
\newblock In {\em NIPS}, pages 2672--2680, 2014.

\bibitem{gulrajani2017improved}
Ishaan Gulrajani, Faruk Ahmed, Martin Arjovsky, Vincent Dumoulin, and Aaron~C
  Courville.
\newblock Improved training of wasserstein gans.
\newblock In {\em NIPS}, pages 5767--5777, 2017.

\bibitem{he2016deep}
Kaiming He, Xiangyu Zhang, Shaoqing Ren, and Jian Sun.
\newblock Deep residual learning for image recognition.
\newblock In {\em CVPR}, pages 770--778, 2016.

\bibitem{huang2015single}
Jia-Bin Huang, Abhishek Singh, and Narendra Ahuja.
\newblock Single image super-resolution from transformed self-exemplars.
\newblock In {\em CVPR}, pages 5197--5206, 2015.

\bibitem{irani1991improving}
Michal Irani and Shmuel Peleg.
\newblock Improving resolution by image registration.
\newblock {\em CVGIP}, 53(3):231--239, 1991.

\bibitem{johnson2016perceptual}
Justin Johnson, Alexandre Alahi, and Li Fei-Fei.
\newblock Perceptual losses for real-time style transfer and super-resolution.
\newblock In {\em ECCV}, pages 694--711. Springer, 2016.

\bibitem{RaGAN}
Alexia Jolicoeur-Martineau.
\newblock The relativistic discriminator: a key element missing from standard
  gan, 2018.

\bibitem{classicalCubic}
Robert Keys.
\newblock Cubic convolution interpolation for digital image processing. ieee
  trans acoust speech signal process.
\newblock {\em TASSP}, 29:1153 -- 1160, 01 1982.

\bibitem{VDSR}
Jiwon Kim, Jung Kwon~Lee, and Kyoung Mu~Lee.
\newblock Accurate image super-resolution using very deep convolutional
  networks.
\newblock In {\em CVPR}, pages 1646--1654, 2016.

\bibitem{DRCN}
Jiwon Kim, Jung Kwon~Lee, and Kyoung Mu~Lee.
\newblock Deeply-recursive convolutional network for image super-resolution.
\newblock In {\em CVPR}, pages 1637--1645, 2016.

\bibitem{kim2010single}
Kwang~In Kim and Younghee Kwon.
\newblock Single-image super-resolution using sparse regression and natural
  image prior.
\newblock {\em TPAMI}, 32(6):1127--1133, 2010.

\bibitem{Adam}
Diederik~P Kingma and Jimmy Ba.
\newblock Adam: A method for stochastic optimization.
\newblock {\em arXiv preprint arXiv:1412.6980}, 2014.

\bibitem{ledig2017photo}
Christian Ledig, Lucas Theis, Ferenc Husz{\'a}r, Jose Caballero, Andrew
  Cunningham, Alejandro Acosta, Andrew Aitken, Alykhan Tejani, Johannes Totz,
  Zehan Wang, et~al.
\newblock Photo-realistic single image super-resolution using a generative
  adversarial network.
\newblock In {\em CVPR}, pages 4681--4690, 2017.

\bibitem{li2019feedback}
Zhen Li, Jinglei Yang, Zheng Liu, Xiaomin Yang, Gwanggil Jeon, and Wei Wu.
\newblock Feedback network for image super-resolution.
\newblock In {\em CVPR}, pages 3867--3876, 2019.

\bibitem{luan2017deep}
Fujun Luan, Sylvain Paris, Eli Shechtman, and Kavita Bala.
\newblock Deep photo style transfer.
\newblock In {\em CVPR}, pages 4990--4998, 2017.

\bibitem{BSD100}
David~R. Martin, Charless~C. Fowlkes, Doron Tal, and Jitendra Malik.
\newblock A database of human segmented natural images and its application to
  evaluating segmentation algorithms and measuring ecological statistics.
\newblock In {\em ICCV}, pages 416--425, 2001.

\bibitem{PyTorch}
Adam Paszke, Sam Gross, Soumith Chintala, Gregory Chanan, Edward Yang, Zachary
  DeVito, Zeming Lin, Alban Desmaison, Luca Antiga, and Adam Lerer.
\newblock Automatic differentiation in pytorch.
\newblock In {\em NIPS-W}, 2017.

\bibitem{rad2019srobb}
Mohammad~Saeed Rad, Behzad Bozorgtabar, Urs-Viktor Marti, Max Basler,
  Hazim~Kemal Ekenel, and Jean-Philippe Thiran.
\newblock Srobb: Targeted perceptual loss for single image super-resolution.
\newblock {\em arXiv preprint arXiv:1908.07222}, 2019.

\bibitem{radford2015unsupervised}
Alec Radford, Luke Metz, and Soumith Chintala.
\newblock Unsupervised representation learning with deep convolutional
  generative adversarial networks.
\newblock {\em arXiv preprint arXiv:1511.06434}, 2015.

\bibitem{EnhanceNet}
Mehdi~SM Sajjadi, Bernhard Scholkopf, and Michael Hirsch.
\newblock Enhancenet: Single image super-resolution through automated texture
  synthesis.
\newblock In {\em ICCV}, pages 4491--4500, 2017.

\bibitem{shi2016real}
Wenzhe Shi, Jose Caballero, Ferenc Husz{\'a}r, Johannes Totz, Andrew~P Aitken,
  Rob Bishop, Daniel Rueckert, and Zehan Wang.
\newblock Real-time single image and video super-resolution using an efficient
  sub-pixel convolutional neural network.
\newblock In {\em CVPR}, pages 1874--1883, 2016.

\bibitem{simonyan2014very}
Karen Simonyan and Andrew Zisserman.
\newblock Very deep convolutional networks for large-scale image recognition.
\newblock {\em arXiv preprint arXiv:1409.1556}, 2014.

\bibitem{soh2019natural}
Jae~Woong Soh, Gu~Yong Park, Junho Jo, and Nam~Ik Cho.
\newblock Natural and realistic single image super-resolution with explicit
  natural manifold discrimination.
\newblock In {\em CVPR}, pages 8122--8131, 2019.

\bibitem{sun2008image}
Jian Sun, Zongben Xu, and Heung-Yeung Shum.
\newblock Image super-resolution using gradient profile prior.
\newblock In {\em CVPR}, pages 1--8. IEEE, 2008.

\bibitem{sun2010gradient}
Jian Sun, Zongben Xu, and Heung-Yeung Shum.
\newblock Gradient profile prior and its applications in image super-resolution
  and enhancement.
\newblock {\em TIP}, 20(6):1529--1542, 2010.

\bibitem{wang2019cfsnet}
Wei Wang, Ruiming Guo, Yapeng Tian, and Wenming Yang.
\newblock Cfsnet: Toward a controllable feature space for image restoration.
\newblock {\em arXiv preprint arXiv:1904.00634}, 2019.

\bibitem{SFTGAN}
Xintao Wang, Ke Yu, Chao Dong, and Chen Change~Loy.
\newblock Recovering realistic texture in image super-resolution by deep
  spatial feature transform.
\newblock In {\em CVPR}, pages 606--615, 2018.

\bibitem{wang2018esrgan}
Xintao Wang, Ke Yu, Shixiang Wu, Jinjin Gu, Yihao Liu, Chao Dong, Yu Qiao, and
  Chen~Change Loy.
\newblock Esrgan: Enhanced super-resolution generative adversarial networks.
\newblock In {\em ECCV}, pages 63--79. Springer, 2018.

\bibitem{wang2004image}
Zhou Wang, Alan~C Bovik, Hamid~R Sheikh, Eero~P Simoncelli, et~al.
\newblock Image quality assessment: from error visibility to structural
  similarity.
\newblock {\em TIP}, 13(4):600--612, 2004.

\bibitem{xiong2010robust}
Zhiwei Xiong, Xiaoyan Sun, and Feng Wu.
\newblock Robust web image/video super-resolution.
\newblock {\em TIP}, 19(8):2017--2028, 2010.

\bibitem{yan2015single}
Qing Yan, Yi Xu, Xiaokang Yang, and Truong~Q Nguyen.
\newblock Single image superresolution based on gradient profile sharpness.
\newblock {\em TIP}, 24(10):3187--3202, 2015.

\bibitem{yang2008image}
Jianchao Yang, John Wright, Thomas Huang, and Yi Ma.
\newblock Image super-resolution as sparse representation of raw image patches.
\newblock In {\em CVPR}, pages 1--8, 2008.

\bibitem{yang2010image}
Jianchao Yang, John Wright, Thomas~S Huang, and Yi Ma.
\newblock Image super-resolution via sparse representation.
\newblock {\em TIP}, 19(11):2861--2873, 2010.

\bibitem{yang2017deep}
Wenhan Yang, Jiashi Feng, Jianchao Yang, Fang Zhao, Jiaying Liu, Zongming Guo,
  and Shuicheng Yan.
\newblock Deep edge guided recurrent residual learning for image
  super-resolution.
\newblock {\em TIP}, 26(12):5895--5907, 2017.

\bibitem{zeyde2010single}
Roman Zeyde, Michael Elad, and Matan Protter.
\newblock On single image scale-up using sparse-representations.
\newblock In {\em ICCS}, pages 711--730. Springer, 2010.

\bibitem{zhang2018unreasonable}
Richard Zhang, Phillip Isola, Alexei~A Efros, Eli Shechtman, and Oliver Wang.
\newblock The unreasonable effectiveness of deep features as a perceptual
  metric.
\newblock In {\em CVPR}, pages 586--595, 2018.

\bibitem{RCAN}
Yulun Zhang, Kunpeng Li, Kai Li, Lichen Wang, Bineng Zhong, and Yun Fu.
\newblock Image super-resolution using very deep residual channel attention
  networks.
\newblock In {\em ECCV}, pages 286--301, 2018.

\bibitem{RDN}
Yulun Zhang, Yapeng Tian, Yu Kong, Bineng Zhong, and Yun Fu.
\newblock Residual dense network for image super-resolution.
\newblock In {\em CVPR}, pages 2472--2481, 2018.

\bibitem{zhu2015modeling}
Yu Zhu, Yanning Zhang, Boyan Bonev, and Alan~L Yuille.
\newblock Modeling deformable gradient compositions for single-image
  super-resolution.
\newblock In {\em CVPR}, 2015.

\end{thebibliography}
}

\appendix

\section*{Supplementary Material}

\section{User Study}

\begin{figure}[b]
\vspace{-5mm}
    \centering
    \includegraphics[width=\linewidth]{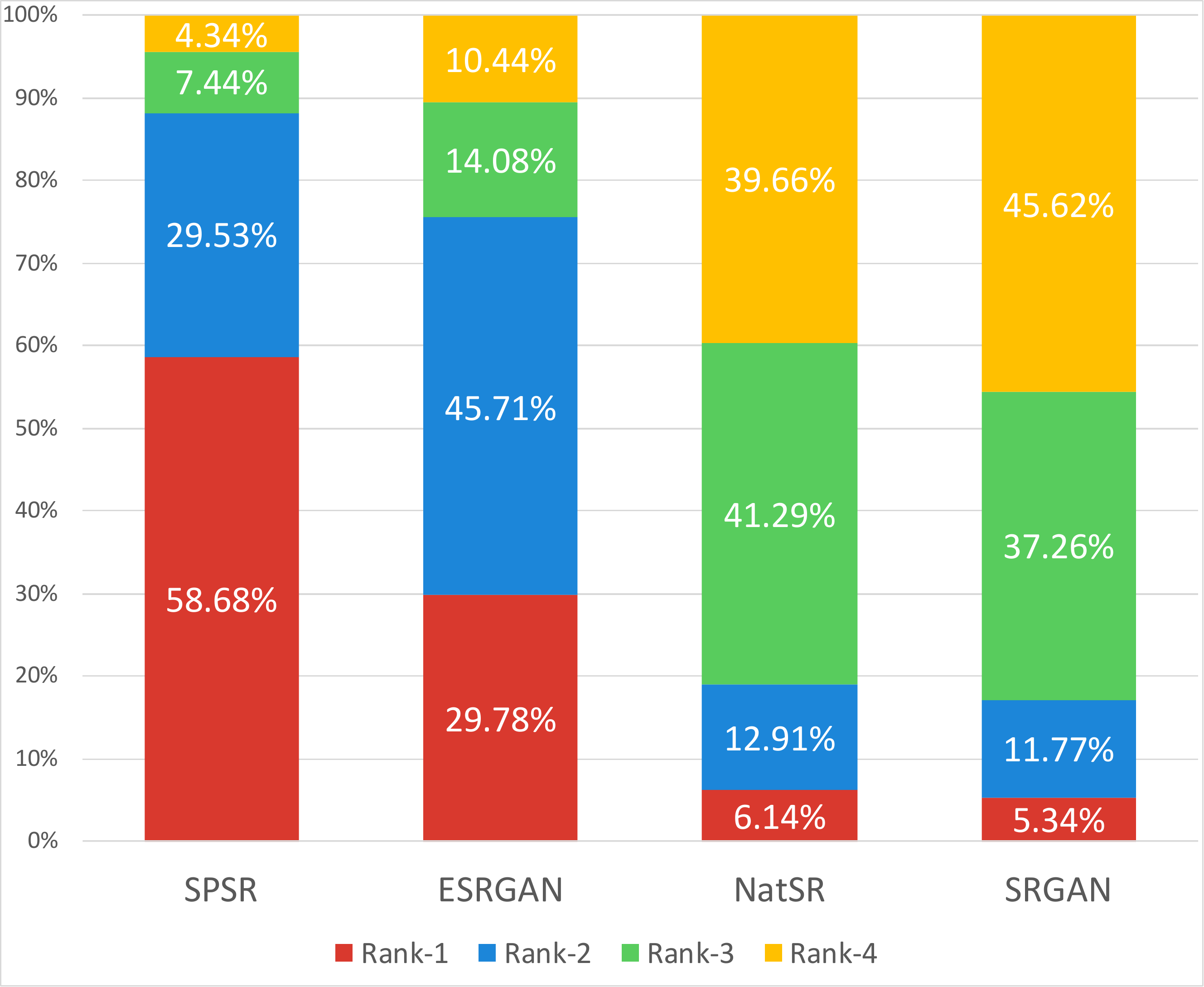}
    \caption{User study results of different GAN-based SR methods. Our SPSR method outperforms state-of-the-art SR methods in generating high-quality images. }
    \label{fig:user-study}
\end{figure}

We conduct a user study as a subjective assessment to evaluate the visual performance of different SR methods on benchmark datasets. HR images are displayed as references while SR results of our SPSR method, ESRGAN~\cite{wang2018esrgan}, NatSR~\cite{soh2019natural} and SRGAN~\cite{ledig2017photo} are presented in a randomized sequence. Human raters are asked to rank the four SR versions according to the perceptual quality. Finally, we collect 1290 votes from 43 human raters. The summarized results are presented in Figure~\ref{fig:user-study}. As shown, our SPSR method gets much more votes of rank-1 than ESRGAN, NatSR and SRGAN. Meanwhile, most SR results of ESRGAN are voted the second best among the four methods since there are more structural distortions in the recovered images of ESRGAN than ours. NatSR and SRGAN fail to obtain satisfactory results. We think the reason is that they sometimes generate relatively blurry textures and undesirable artifacts. The comparison with the state-of-the-art GAN-based SR methods verifies the superiority of our proposed method in generating high-fidelity SR results. 

\section{More Qualitative Results}

\begin{figure*}[htbp]
\centering

\begin{minipage}[b]{\picwidth\linewidth}
\includegraphics[width=1 \linewidth]{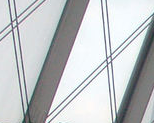}\vspace{\picvspace}
\centering{\scriptsize{HR ('img\_002' from Urban100)}}
\end{minipage}\hspace{\pagehspace}
\begin{minipage}[b]{\picwidth\linewidth}
\includegraphics[width=1 \linewidth]{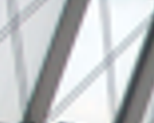}\vspace{\picvspace}
\centering{\scriptsize{LR}}
\end{minipage}\hspace{\pagehspace}
\begin{minipage}[b]{\picwidth\linewidth}
\includegraphics[width=1 \linewidth]{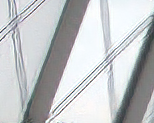}\vspace{\picvspace}
\centering{\scriptsize{EnhanceNet}}
\end{minipage}\hspace{\pagehspace}
\begin{minipage}[b]{\picwidth\linewidth}
\includegraphics[width=1 \linewidth]{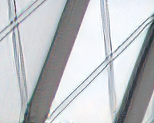}\vspace{\picvspace}
\centering{\scriptsize{SFTGAN}}
\end{minipage}\vspace{\pagevspace}

\begin{minipage}[b]{\picwidth\linewidth}
\includegraphics[width=1 \linewidth]{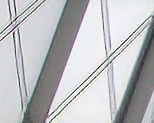}\vspace{\picvspace}
\centering{\scriptsize{SRGAN}}
\end{minipage}\hspace{\pagehspace}
\begin{minipage}[b]{\picwidth\linewidth}
\includegraphics[width=1 \linewidth]{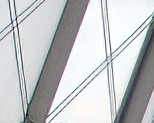}\vspace{\picvspace}
\centering{\scriptsize{ESRGAN}}
\end{minipage}\hspace{\pagehspace}
\begin{minipage}[b]{\picwidth\linewidth}
\includegraphics[width=1 \linewidth]{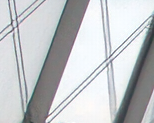}\vspace{\picvspace}
\centering{\scriptsize{NatSR}}
\end{minipage}\hspace{\pagehspace}
\begin{minipage}[b]{\picwidth\linewidth}
\includegraphics[width=1 \linewidth]{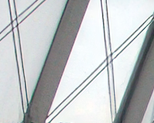}\vspace{\picvspace}
\centering{\scriptsize{SPSR}}
\end{minipage}\vspace{\packvspace}

\begin{minipage}[b]{\picwidth\linewidth}
\includegraphics[width=1 \linewidth]{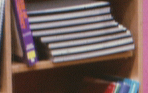}\vspace{\picvspace}
\centering{\scriptsize{HR ('barbara' from Set14)}}
\end{minipage}\hspace{\pagehspace}
\begin{minipage}[b]{\picwidth\linewidth}
\includegraphics[width=1 \linewidth]{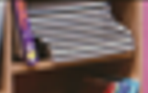}\vspace{\picvspace}
\centering{\scriptsize{LR}}
\end{minipage}\hspace{\pagehspace}
\begin{minipage}[b]{\picwidth\linewidth}
\includegraphics[width=1 \linewidth]{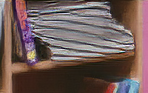}\vspace{\picvspace}
\centering{\scriptsize{EnhanceNet}}
\end{minipage}\hspace{\pagehspace}
\begin{minipage}[b]{\picwidth\linewidth}
\includegraphics[width=1 \linewidth]{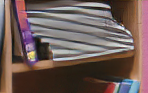}\vspace{\picvspace}
\centering{\scriptsize{SFTGAN}}
\end{minipage}\vspace{\pagevspace}

\begin{minipage}[b]{\picwidth\linewidth}
\includegraphics[width=1 \linewidth]{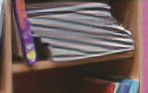}\vspace{\picvspace}
\centering{\scriptsize{SRGAN}}
\end{minipage}\hspace{\pagehspace}
\begin{minipage}[b]{\picwidth\linewidth}
\includegraphics[width=1 \linewidth]{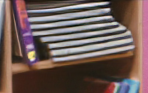}\vspace{\picvspace}
\centering{\scriptsize{ESRGAN}}
\end{minipage}\hspace{\pagehspace}
\begin{minipage}[b]{\picwidth\linewidth}
\includegraphics[width=1 \linewidth]{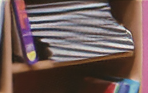}\vspace{\picvspace}
\centering{\scriptsize{NatSR}}
\end{minipage}\hspace{\pagehspace}
\begin{minipage}[b]{\picwidth\linewidth}
\includegraphics[width=1 \linewidth]{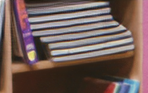}\vspace{\picvspace}
\centering{\scriptsize{SPSR}}
\end{minipage}\vspace{\packvspace}

\begin{minipage}[b]{\picwidth\linewidth}
\includegraphics[width=1 \linewidth]{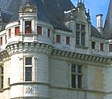}\vspace{\picvspace}
\centering{\scriptsize{HR ('102061' from BSD100)}}
\end{minipage}\hspace{\pagehspace}
\begin{minipage}[b]{\picwidth\linewidth}
\includegraphics[width=1 \linewidth]{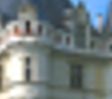}\vspace{\picvspace}
\centering{\scriptsize{LR}}
\end{minipage}\hspace{\pagehspace}
\begin{minipage}[b]{\picwidth\linewidth}
\includegraphics[width=1 \linewidth]{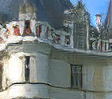}\vspace{\picvspace}
\centering{\scriptsize{EnhanceNet}}
\end{minipage}\hspace{\pagehspace}
\begin{minipage}[b]{\picwidth\linewidth}
\includegraphics[width=1 \linewidth]{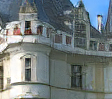}\vspace{\picvspace}
\centering{\scriptsize{SFTGAN}}
\end{minipage}\vspace{\pagevspace}

\begin{minipage}[b]{\picwidth\linewidth}
\includegraphics[width=1 \linewidth]{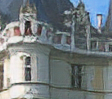}\vspace{\picvspace}
\centering{\scriptsize{SRGAN}}
\end{minipage}\hspace{\pagehspace}
\begin{minipage}[b]{\picwidth\linewidth}
\includegraphics[width=1 \linewidth]{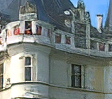}\vspace{\picvspace}
\centering{\scriptsize{ESRGAN}}
\end{minipage}\hspace{\pagehspace}
\begin{minipage}[b]{\picwidth\linewidth}
\includegraphics[width=1 \linewidth]{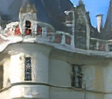}\vspace{\picvspace}
\centering{\scriptsize{NatSR}}
\end{minipage}\hspace{\pagehspace}
\begin{minipage}[b]{\picwidth\linewidth}
\includegraphics[width=1 \linewidth]{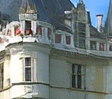}\vspace{\picvspace}
\centering{\scriptsize{SPSR}}
\end{minipage}\vspace{\captionvspace}

\centering
\caption{Visual comparison of SR performance with state-of-the-art SR methods.}
\label{fig:supp_vis1}
\end{figure*}
\begin{figure*}[htbp]
\centering

\begin{minipage}[b]{\picwidth\linewidth}
\includegraphics[width=1 \linewidth]{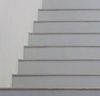}\vspace{\picvspace}
\centering{\scriptsize{HR ('img\_009' from Urban100)}}
\end{minipage}\hspace{\pagehspace}
\begin{minipage}[b]{\picwidth\linewidth}
\includegraphics[width=1 \linewidth]{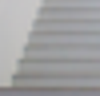}\vspace{\picvspace}
\centering{\scriptsize{LR}}
\end{minipage}\hspace{\pagehspace}
\begin{minipage}[b]{\picwidth\linewidth}
\includegraphics[width=1 \linewidth]{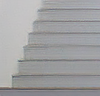}\vspace{\picvspace}
\centering{\scriptsize{EnhanceNet}}
\end{minipage}\hspace{\pagehspace}
\begin{minipage}[b]{\picwidth\linewidth}
\includegraphics[width=1 \linewidth]{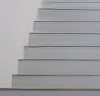}\vspace{\picvspace}
\centering{\scriptsize{SFTGAN}}
\end{minipage}\vspace{\pagevspace}

\begin{minipage}[b]{\picwidth\linewidth}
\includegraphics[width=1 \linewidth]{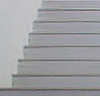}\vspace{\picvspace}
\centering{\scriptsize{SRGAN}}
\end{minipage}\hspace{\pagehspace}
\begin{minipage}[b]{\picwidth\linewidth}
\includegraphics[width=1 \linewidth]{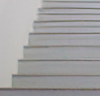}\vspace{\picvspace}
\centering{\scriptsize{ESRGAN}}
\end{minipage}\hspace{\pagehspace}
\begin{minipage}[b]{\picwidth\linewidth}
\includegraphics[width=1 \linewidth]{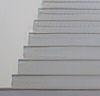}\vspace{\picvspace}
\centering{\scriptsize{NatSR}}
\end{minipage}\hspace{\pagehspace}
\begin{minipage}[b]{\picwidth\linewidth}
\includegraphics[width=1 \linewidth]{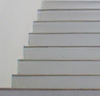}\vspace{\picvspace}
\centering{\scriptsize{SPSR}}
\end{minipage}\vspace{\packvspace}

\begin{minipage}[b]{\picwidth\linewidth}
\includegraphics[width=1 \linewidth]{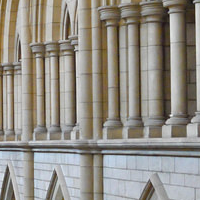}\vspace{\picvspace}
\centering{\scriptsize{HR ('img\_065' from Urban100)}}
\end{minipage}\hspace{\pagehspace}
\begin{minipage}[b]{\picwidth\linewidth}
\includegraphics[width=1 \linewidth]{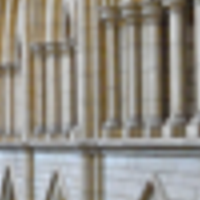}\vspace{\picvspace}
\centering{\scriptsize{LR}}
\end{minipage}\hspace{\pagehspace}
\begin{minipage}[b]{\picwidth\linewidth}
\includegraphics[width=1 \linewidth]{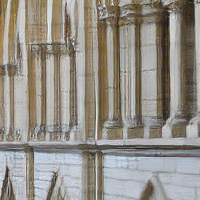}\vspace{\picvspace}
\centering{\scriptsize{EnhanceNet}}
\end{minipage}\hspace{\pagehspace}
\begin{minipage}[b]{\picwidth\linewidth}
\includegraphics[width=1 \linewidth]{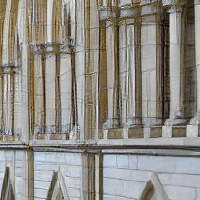}\vspace{\picvspace}
\centering{\scriptsize{SFTGAN}}
\end{minipage}\vspace{\pagevspace}

\begin{minipage}[b]{\picwidth\linewidth}
\includegraphics[width=1 \linewidth]{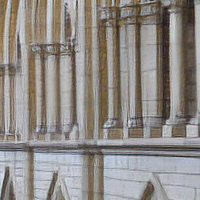}\vspace{\picvspace}
\centering{\scriptsize{SRGAN}}
\end{minipage}\hspace{\pagehspace}
\begin{minipage}[b]{\picwidth\linewidth}
\includegraphics[width=1 \linewidth]{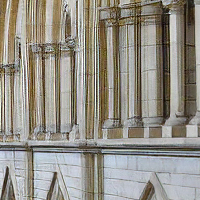}\vspace{\picvspace}
\centering{\scriptsize{ESRGAN}}
\end{minipage}\hspace{\pagehspace}
\begin{minipage}[b]{\picwidth\linewidth}
\includegraphics[width=1 \linewidth]{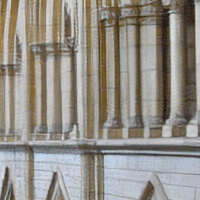}\vspace{\picvspace}
\centering{\scriptsize{NatSR}}
\end{minipage}\hspace{\pagehspace}
\begin{minipage}[b]{\picwidth\linewidth}
\includegraphics[width=1 \linewidth]{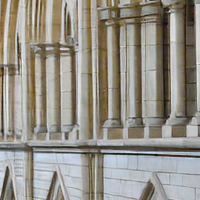}\vspace{\picvspace}
\centering{\scriptsize{SPSR}}
\end{minipage}\vspace{\packvspace}

\begin{minipage}[b]{\picwidth\linewidth}
\includegraphics[width=1 \linewidth]{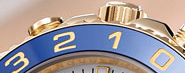}\vspace{\picvspace}
\centering{\scriptsize{HR ('im\_024' from General100)}}
\end{minipage}\hspace{\pagehspace}
\begin{minipage}[b]{\picwidth\linewidth}
\includegraphics[width=1 \linewidth]{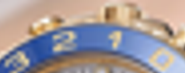}\vspace{\picvspace}
\centering{\scriptsize{LR}}
\end{minipage}\hspace{\pagehspace}
\begin{minipage}[b]{\picwidth\linewidth}
\includegraphics[width=1 \linewidth]{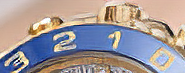}\vspace{\picvspace}
\centering{\scriptsize{EnhanceNet}}
\end{minipage}\hspace{\pagehspace}
\begin{minipage}[b]{\picwidth\linewidth}
\includegraphics[width=1 \linewidth]{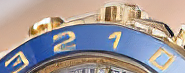}\vspace{\picvspace}
\centering{\scriptsize{SFTGAN}}
\end{minipage}\vspace{\pagevspace}

\begin{minipage}[b]{\picwidth\linewidth}
\includegraphics[width=1 \linewidth]{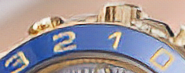}\vspace{\picvspace}
\centering{\scriptsize{SRGAN}}
\end{minipage}\hspace{\pagehspace}
\begin{minipage}[b]{\picwidth\linewidth}
\includegraphics[width=1 \linewidth]{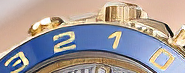}\vspace{\picvspace}
\centering{\scriptsize{ESRGAN}}
\end{minipage}\hspace{\pagehspace}
\begin{minipage}[b]{\picwidth\linewidth}
\includegraphics[width=1 \linewidth]{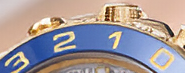}\vspace{\picvspace}
\centering{\scriptsize{NatSR}}
\end{minipage}\hspace{\pagehspace}
\begin{minipage}[b]{\picwidth\linewidth}
\includegraphics[width=1 \linewidth]{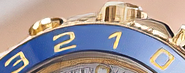}\vspace{\picvspace}
\centering{\scriptsize{SPSR}}
\end{minipage}\vspace{\captionvspace}

\centering
\caption{Visual comparison of SR performance with state-of-the-art SR methods.}
\label{fig:supp_vis2}
\end{figure*}
\begin{figure*}[htbp]
\centering

\begin{minipage}[b]{\picwidth\linewidth}
\includegraphics[width=1 \linewidth]{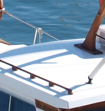}\vspace{\picvspace}
\centering{\scriptsize{HR ('im\_023' from General100)}}
\end{minipage}\hspace{\pagehspace}
\begin{minipage}[b]{\picwidth\linewidth}
\includegraphics[width=1 \linewidth]{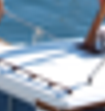}\vspace{\picvspace}
\centering{\scriptsize{LR}}
\end{minipage}\hspace{\pagehspace}
\begin{minipage}[b]{\picwidth\linewidth}
\includegraphics[width=1 \linewidth]{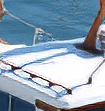}\vspace{\picvspace}
\centering{\scriptsize{EnhanceNet}}
\end{minipage}\hspace{\pagehspace}
\begin{minipage}[b]{\picwidth\linewidth}
\includegraphics[width=1 \linewidth]{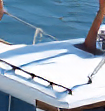}\vspace{\picvspace}
\centering{\scriptsize{SFTGAN}}
\end{minipage}\vspace{\pagevspace}

\begin{minipage}[b]{\picwidth\linewidth}
\includegraphics[width=1 \linewidth]{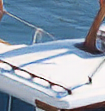}\vspace{\picvspace}
\centering{\scriptsize{SRGAN}}
\end{minipage}\hspace{\pagehspace}
\begin{minipage}[b]{\picwidth\linewidth}
\includegraphics[width=1 \linewidth]{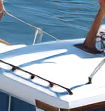}\vspace{\picvspace}
\centering{\scriptsize{ESRGAN}}
\end{minipage}\hspace{\pagehspace}
\begin{minipage}[b]{\picwidth\linewidth}
\includegraphics[width=1 \linewidth]{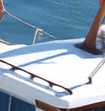}\vspace{\picvspace}
\centering{\scriptsize{NatSR}}
\end{minipage}\hspace{\pagehspace}
\begin{minipage}[b]{\picwidth\linewidth}
\includegraphics[width=1 \linewidth]{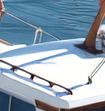}\vspace{\picvspace}
\centering{\scriptsize{SPSR}}
\end{minipage}\vspace{\packvspace}

\begin{minipage}[b]{\picwidth\linewidth}
\includegraphics[width=1 \linewidth]{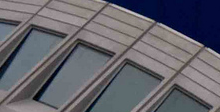}\vspace{\picvspace}
\centering{\scriptsize{HR ('im\_068' from Urban100)}}
\end{minipage}\hspace{\pagehspace}
\begin{minipage}[b]{\picwidth\linewidth}
\includegraphics[width=1 \linewidth]{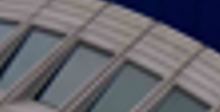}\vspace{\picvspace}
\centering{\scriptsize{LR}}
\end{minipage}\hspace{\pagehspace}
\begin{minipage}[b]{\picwidth\linewidth}
\includegraphics[width=1 \linewidth]{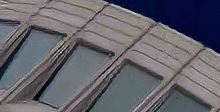}\vspace{\picvspace}
\centering{\scriptsize{EnhanceNet}}
\end{minipage}\hspace{\pagehspace}
\begin{minipage}[b]{\picwidth\linewidth}
\includegraphics[width=1 \linewidth]{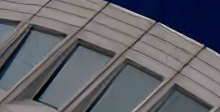}\vspace{\picvspace}
\centering{\scriptsize{SFTGAN}}
\end{minipage}\vspace{\pagevspace}

\begin{minipage}[b]{\picwidth\linewidth}
\includegraphics[width=1 \linewidth]{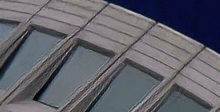}\vspace{\picvspace}
\centering{\scriptsize{SRGAN}}
\end{minipage}\hspace{\pagehspace}
\begin{minipage}[b]{\picwidth\linewidth}
\includegraphics[width=1 \linewidth]{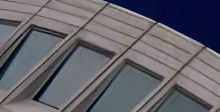}\vspace{\picvspace}
\centering{\scriptsize{ESRGAN}}
\end{minipage}\hspace{\pagehspace}
\begin{minipage}[b]{\picwidth\linewidth}
\includegraphics[width=1 \linewidth]{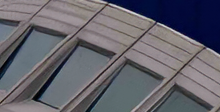}\vspace{\picvspace}
\centering{\scriptsize{NatSR}}
\end{minipage}\hspace{\pagehspace}
\begin{minipage}[b]{\picwidth\linewidth}
\includegraphics[width=1 \linewidth]{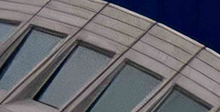}\vspace{\picvspace}
\centering{\scriptsize{SPSR}}
\end{minipage}\vspace{\packvspace}

\begin{minipage}[b]{\picwidth\linewidth}
\includegraphics[width=1 \linewidth]{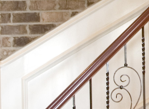}\vspace{\picvspace}
\centering{\scriptsize{HR ('im\_005' from General100)}}
\end{minipage}\hspace{\pagehspace}
\begin{minipage}[b]{\picwidth\linewidth}
\includegraphics[width=1 \linewidth]{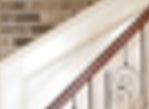}\vspace{\picvspace}
\centering{\scriptsize{LR}}
\end{minipage}\hspace{\pagehspace}
\begin{minipage}[b]{\picwidth\linewidth}
\includegraphics[width=1 \linewidth]{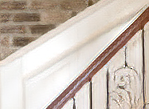}\vspace{\picvspace}
\centering{\scriptsize{EnhanceNet}}
\end{minipage}\hspace{\pagehspace}
\begin{minipage}[b]{\picwidth\linewidth}
\includegraphics[width=1 \linewidth]{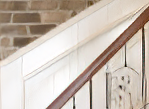}\vspace{\picvspace}
\centering{\scriptsize{SFTGAN}}
\end{minipage}\vspace{\pagevspace}

\begin{minipage}[b]{\picwidth\linewidth}
\includegraphics[width=1 \linewidth]{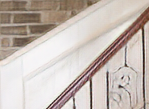}\vspace{\picvspace}
\centering{\scriptsize{SRGAN}}
\end{minipage}\hspace{\pagehspace}
\begin{minipage}[b]{\picwidth\linewidth}
\includegraphics[width=1 \linewidth]{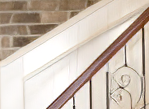}\vspace{\picvspace}
\centering{\scriptsize{ESRGAN}}
\end{minipage}\hspace{\pagehspace}
\begin{minipage}[b]{\picwidth\linewidth}
\includegraphics[width=1 \linewidth]{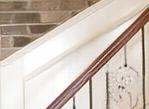}\vspace{\picvspace}
\centering{\scriptsize{NatSR}}
\end{minipage}\hspace{\pagehspace}
\begin{minipage}[b]{\picwidth\linewidth}
\includegraphics[width=1 \linewidth]{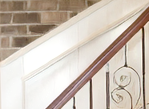}\vspace{\picvspace}
\centering{\scriptsize{SPSR}}
\end{minipage}\vspace{\captionvspace}

\centering
\caption{Visual comparison of SR performance with state-of-the-art SR methods.}
\label{fig:supp_vis3}
\end{figure*}
\begin{figure*}[htbp]
\centering

\begin{minipage}[b]{\picwidth\linewidth}
\includegraphics[width=1 \linewidth]{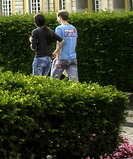}\vspace{\picvspace}
\centering{\scriptsize{HR ('img\_077' from Urban100)}}
\end{minipage}\hspace{\pagehspace}
\begin{minipage}[b]{\picwidth\linewidth}
\includegraphics[width=1 \linewidth]{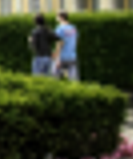}\vspace{\picvspace}
\centering{\scriptsize{LR}}
\end{minipage}\hspace{\pagehspace}
\begin{minipage}[b]{\picwidth\linewidth}
\includegraphics[width=1 \linewidth]{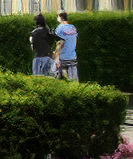}\vspace{\picvspace}
\centering{\scriptsize{EnhanceNet}}
\end{minipage}\hspace{\pagehspace}
\begin{minipage}[b]{\picwidth\linewidth}
\includegraphics[width=1 \linewidth]{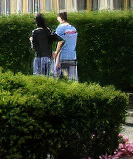}\vspace{\picvspace}
\centering{\scriptsize{SFTGAN}}
\end{minipage}\vspace{\pagevspace}

\begin{minipage}[b]{\picwidth\linewidth}
\includegraphics[width=1 \linewidth]{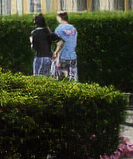}\vspace{\picvspace}
\centering{\scriptsize{SRGAN}}
\end{minipage}\hspace{\pagehspace}
\begin{minipage}[b]{\picwidth\linewidth}
\includegraphics[width=1 \linewidth]{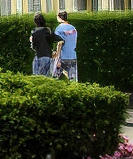}\vspace{\picvspace}
\centering{\scriptsize{ESRGAN}}
\end{minipage}\hspace{\pagehspace}
\begin{minipage}[b]{\picwidth\linewidth}
\includegraphics[width=1 \linewidth]{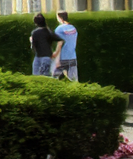}\vspace{\picvspace}
\centering{\scriptsize{NatSR}}
\end{minipage}\hspace{\pagehspace}
\begin{minipage}[b]{\picwidth\linewidth}
\includegraphics[width=1 \linewidth]{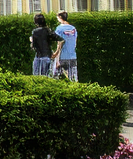}\vspace{\picvspace}
\centering{\scriptsize{SPSR}}
\end{minipage}\vspace{\packvspace}

\begin{minipage}[b]{\picwidth\linewidth}
\includegraphics[width=1 \linewidth]{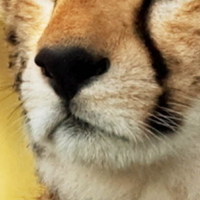}\vspace{\picvspace}
\centering{\scriptsize{HR ('im\_008' from General100)}}
\end{minipage}\hspace{\pagehspace}
\begin{minipage}[b]{\picwidth\linewidth}
\includegraphics[width=1 \linewidth]{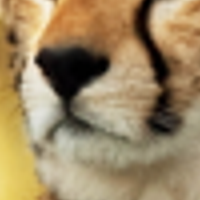}\vspace{\picvspace}
\centering{\scriptsize{LR}}
\end{minipage}\hspace{\pagehspace}
\begin{minipage}[b]{\picwidth\linewidth}
\includegraphics[width=1 \linewidth]{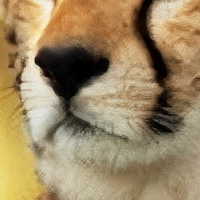}\vspace{\picvspace}
\centering{\scriptsize{EnhanceNet}}
\end{minipage}\hspace{\pagehspace}
\begin{minipage}[b]{\picwidth\linewidth}
\includegraphics[width=1 \linewidth]{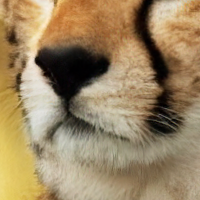}\vspace{\picvspace}
\centering{\scriptsize{SFTGAN}}
\end{minipage}\vspace{\pagevspace}

\begin{minipage}[b]{\picwidth\linewidth}
\includegraphics[width=1 \linewidth]{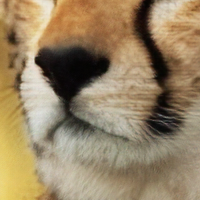}\vspace{\picvspace}
\centering{\scriptsize{SRGAN}}
\end{minipage}\hspace{\pagehspace}
\begin{minipage}[b]{\picwidth\linewidth}
\includegraphics[width=1 \linewidth]{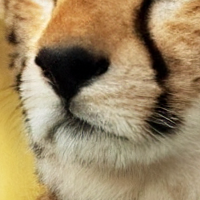}\vspace{\picvspace}
\centering{\scriptsize{ESRGAN}}
\end{minipage}\hspace{\pagehspace}
\begin{minipage}[b]{\picwidth\linewidth}
\includegraphics[width=1 \linewidth]{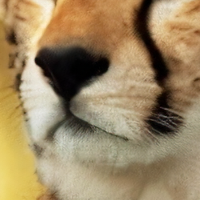}\vspace{\picvspace}
\centering{\scriptsize{NatSR}}
\end{minipage}\hspace{\pagehspace}
\begin{minipage}[b]{\picwidth\linewidth}
\includegraphics[width=1 \linewidth]{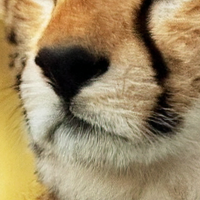}\vspace{\picvspace}
\centering{\scriptsize{SPSR}}
\end{minipage}\vspace{\captionvspace}

\centering
\caption{Visual comparison of SR performance with state-of-the-art SR methods.}
\label{fig:supp_vis4}
\end{figure*}
\begin{figure*}[htbp]

\begin{minipage}[b]{\picwidth\linewidth}
\includegraphics[width=1 \linewidth]{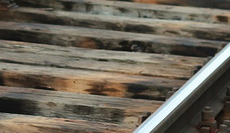}\vspace{\picvspace}
\centering{\scriptsize{HR ('img\_007' from Urban100)}}
\end{minipage}\hspace{\pagehspace}
\begin{minipage}[b]{\picwidth\linewidth}
\includegraphics[width=1 \linewidth]{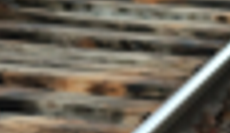}\vspace{\picvspace}
\centering{\scriptsize{LR}}
\end{minipage}\hspace{\pagehspace}
\begin{minipage}[b]{\picwidth\linewidth}
\includegraphics[width=1 \linewidth]{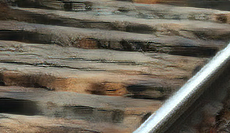}\vspace{\picvspace}
\centering{\scriptsize{EnhanceNet}}
\end{minipage}\hspace{\pagehspace}
\begin{minipage}[b]{\picwidth\linewidth}
\includegraphics[width=1 \linewidth]{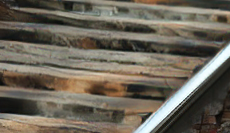}\vspace{\picvspace}
\centering{\scriptsize{SFTGAN}}
\end{minipage}\vspace{\pagevspace}

\begin{minipage}[b]{\picwidth\linewidth}
\includegraphics[width=1 \linewidth]{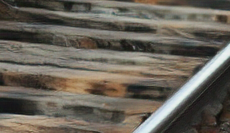}\vspace{\picvspace}
\centering{\scriptsize{SRGAN}}
\end{minipage}\hspace{\pagehspace}
\begin{minipage}[b]{\picwidth\linewidth}
\includegraphics[width=1 \linewidth]{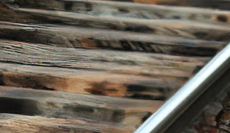}\vspace{\picvspace}
\centering{\scriptsize{ESRGAN}}
\end{minipage}\hspace{\pagehspace}
\begin{minipage}[b]{\picwidth\linewidth}
\includegraphics[width=1 \linewidth]{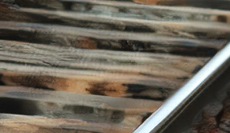}\vspace{\picvspace}
\centering{\scriptsize{NatSR}}
\end{minipage}\hspace{\pagehspace}
\begin{minipage}[b]{\picwidth\linewidth}
\includegraphics[width=1 \linewidth]{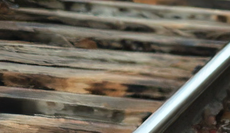}\vspace{\picvspace}
\centering{\scriptsize{SPSR}}
\end{minipage}\vspace{\packvspace}

\begin{minipage}[b]{\picwidth\linewidth}
\includegraphics[width=1 \linewidth]{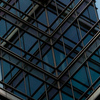}\vspace{\picvspace}
\centering{\scriptsize{HR ('img\_047' from Urban100)}}
\end{minipage}\hspace{\pagehspace}
\begin{minipage}[b]{\picwidth\linewidth}
\includegraphics[width=1 \linewidth]{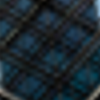}\vspace{\picvspace}
\centering{\scriptsize{LR}}
\end{minipage}\hspace{\pagehspace}
\begin{minipage}[b]{\picwidth\linewidth}
\includegraphics[width=1 \linewidth]{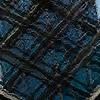}\vspace{\picvspace}
\centering{\scriptsize{EnhanceNet}}
\end{minipage}\hspace{\pagehspace}
\begin{minipage}[b]{\picwidth\linewidth}
\includegraphics[width=1 \linewidth]{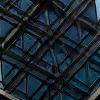}\vspace{\picvspace}
\centering{\scriptsize{SFTGAN}}
\end{minipage}\vspace{\pagevspace}

\begin{minipage}[b]{\picwidth\linewidth}
\includegraphics[width=1 \linewidth]{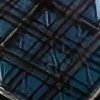}\vspace{\picvspace}
\centering{\scriptsize{SRGAN}}
\end{minipage}\hspace{\pagehspace}
\begin{minipage}[b]{\picwidth\linewidth}
\includegraphics[width=1 \linewidth]{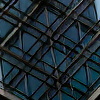}\vspace{\picvspace}
\centering{\scriptsize{ESRGAN}}
\end{minipage}\hspace{\pagehspace}
\begin{minipage}[b]{\picwidth\linewidth}
\includegraphics[width=1 \linewidth]{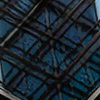}\vspace{\picvspace}
\centering{\scriptsize{NatSR}}
\end{minipage}\hspace{\pagehspace}
\begin{minipage}[b]{\picwidth\linewidth}
\includegraphics[width=1 \linewidth]{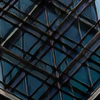}\vspace{\picvspace}
\centering{\scriptsize{SPSR}}
\end{minipage}\vspace{\packvspace}

\begin{minipage}[b]{\picwidth\linewidth}
\includegraphics[width=1 \linewidth]{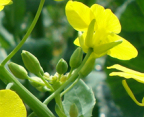}\vspace{\picvspace}
\centering{\scriptsize{HR ('im\_090' from General100)}}
\end{minipage}\hspace{\pagehspace}
\begin{minipage}[b]{\picwidth\linewidth}
\includegraphics[width=1 \linewidth]{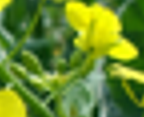}\vspace{\picvspace}
\centering{\scriptsize{LR}}
\end{minipage}\hspace{\pagehspace}
\begin{minipage}[b]{\picwidth\linewidth}
\includegraphics[width=1 \linewidth]{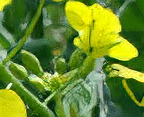}\vspace{\picvspace}
\centering{\scriptsize{EnhanceNet}}
\end{minipage}\hspace{\pagehspace}
\begin{minipage}[b]{\picwidth\linewidth}
\includegraphics[width=1 \linewidth]{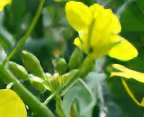}\vspace{\picvspace}
\centering{\scriptsize{SFTGAN}}
\end{minipage}\vspace{\pagevspace}

\begin{minipage}[b]{\picwidth\linewidth}
\includegraphics[width=1 \linewidth]{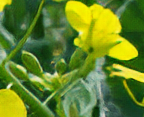}\vspace{\picvspace}
\centering{\scriptsize{SRGAN}}
\end{minipage}\hspace{\pagehspace}
\begin{minipage}[b]{\picwidth\linewidth}
\includegraphics[width=1 \linewidth]{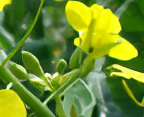}\vspace{\picvspace}
\centering{\scriptsize{ESRGAN}}
\end{minipage}\hspace{\pagehspace}
\begin{minipage}[b]{\picwidth\linewidth}
\includegraphics[width=1 \linewidth]{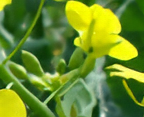}\vspace{\picvspace}
\centering{\scriptsize{NatSR}}
\end{minipage}\hspace{\pagehspace}
\begin{minipage}[b]{\picwidth\linewidth}
\includegraphics[width=1 \linewidth]{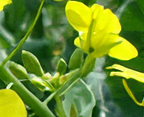}\vspace{\picvspace}
\centering{\scriptsize{SPSR}}
\end{minipage}\vspace{\captionvspace}

\centering

\centering
\caption{Visual comparison of SR performance with state-of-the-art SR methods.}
\label{fig:supp_vis5}
\end{figure*}
\begin{figure*}[htbp]
\centering

\begin{minipage}[b]{\picwidth\linewidth}
\includegraphics[width=1 \linewidth]{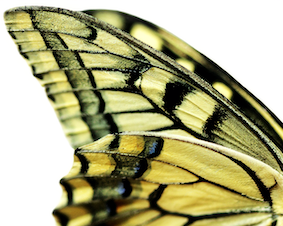}\vspace{\picvspace}
\centering{\scriptsize{HR ('im\_014' from General100)}}
\end{minipage}\hspace{\pagehspace}
\begin{minipage}[b]{\picwidth\linewidth}
\includegraphics[width=1 \linewidth]{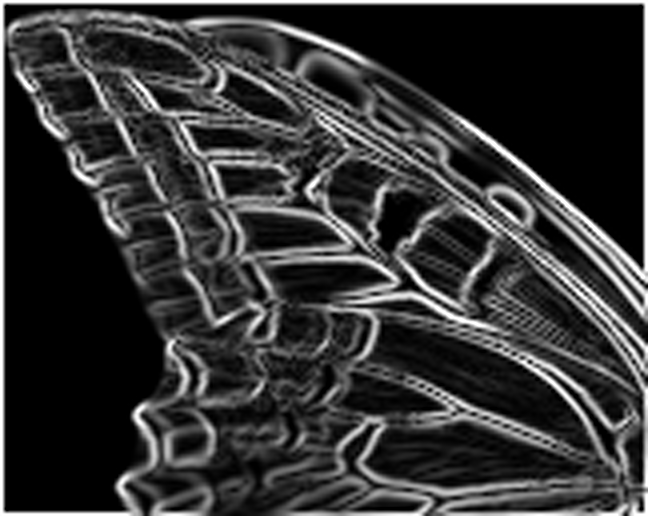}\vspace{\picvspace}
\centering{\scriptsize{LR gradient (Bicubic)}}
\end{minipage}\hspace{\pagehspace}
\begin{minipage}[b]{\picwidth\linewidth}
\includegraphics[width=1 \linewidth]{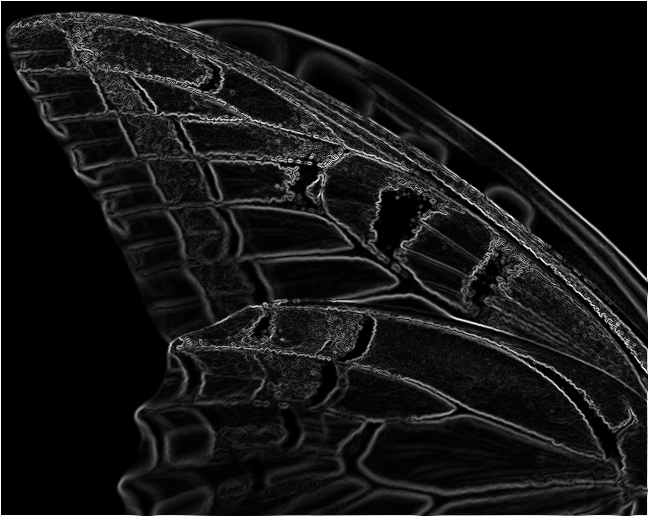}\vspace{\picvspace}
\centering{\scriptsize{HR gradient}}
\end{minipage}\hspace{\pagehspace}
\begin{minipage}[b]{\picwidth\linewidth}
\includegraphics[width=1 \linewidth]{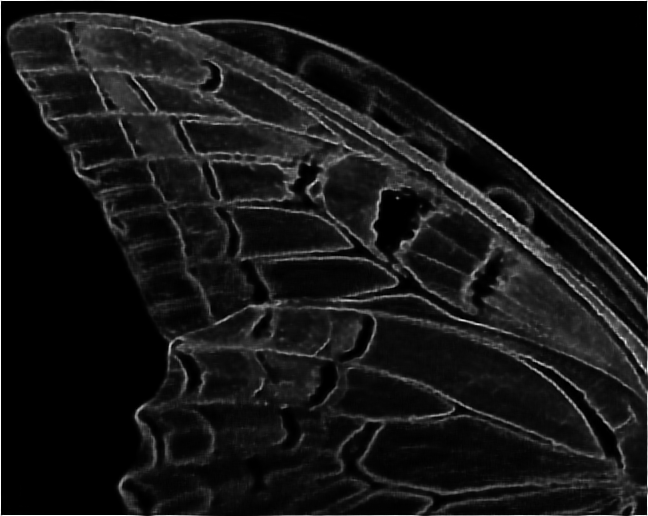}\vspace{\picvspace}
\centering{\scriptsize{Output of the gradient branch}}
\end{minipage}\vspace{\packvspace}

\begin{minipage}[b]{\picwidth\linewidth}
\includegraphics[width=1 \linewidth]{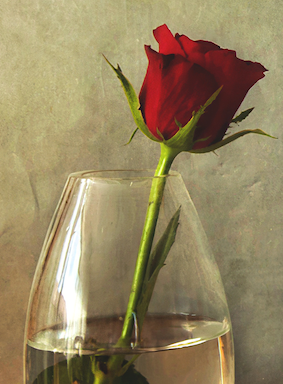}\vspace{\picvspace}
\centering{\scriptsize{HR ('im\_026' from General100)}}
\end{minipage}\hspace{\pagehspace}
\begin{minipage}[b]{\picwidth\linewidth}
\includegraphics[width=1 \linewidth]{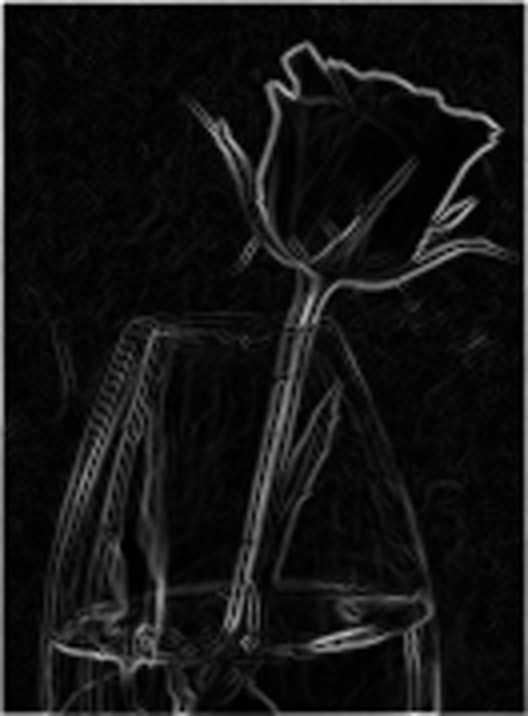}\vspace{\picvspace}
\centering{\scriptsize{LR gradient (Bicubic)}}
\end{minipage}\hspace{\pagehspace}
\begin{minipage}[b]{\picwidth\linewidth}
\includegraphics[width=1 \linewidth]{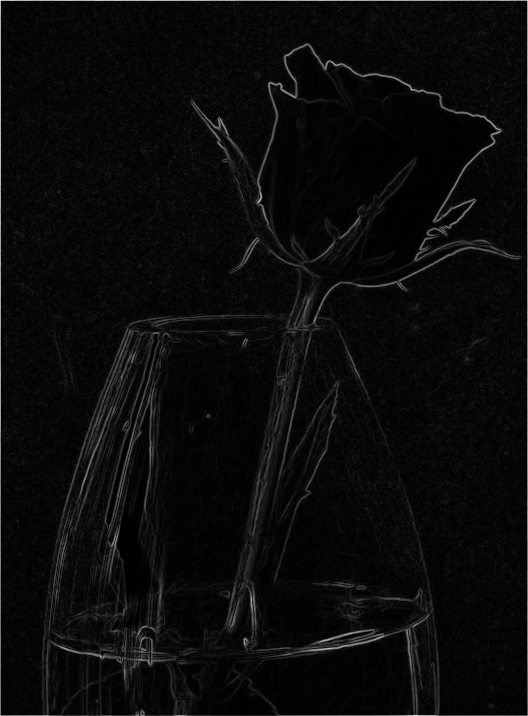}\vspace{\picvspace}
\centering{\scriptsize{HR gradient}}
\end{minipage}\hspace{\pagehspace}
\begin{minipage}[b]{\picwidth\linewidth}
\includegraphics[width=1 \linewidth]{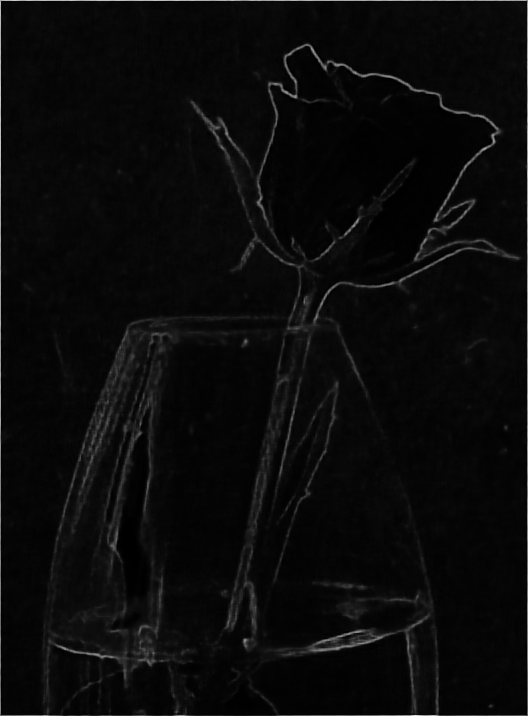}\vspace{\picvspace}
\centering{\scriptsize{Output of the gradient branch}}
\end{minipage}\vspace{\packvspace}

\begin{minipage}[b]{\picwidth\linewidth}
\includegraphics[width=1 \linewidth]{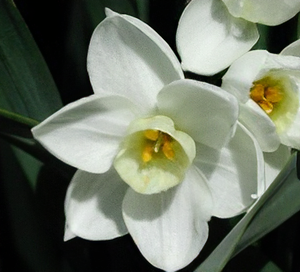}\vspace{\picvspace}
\centering{\scriptsize{HR ('im\_055' from General100)}}
\end{minipage}\hspace{\pagehspace}
\begin{minipage}[b]{\picwidth\linewidth}
\includegraphics[width=1 \linewidth]{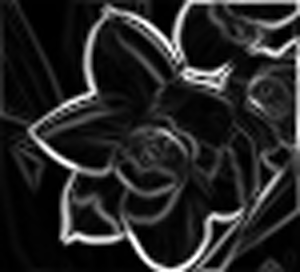}\vspace{\picvspace}
\centering{\scriptsize{LR gradient (Bicubic)}}
\end{minipage}\hspace{\pagehspace}
\begin{minipage}[b]{\picwidth\linewidth}
\includegraphics[width=1 \linewidth]{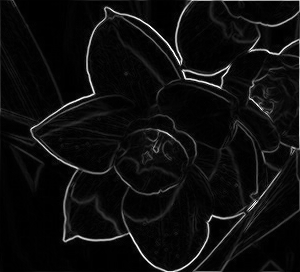}\vspace{\picvspace}
\centering{\scriptsize{HR gradient}}
\end{minipage}\hspace{\pagehspace}
\begin{minipage}[b]{\picwidth\linewidth}
\includegraphics[width=1 \linewidth]{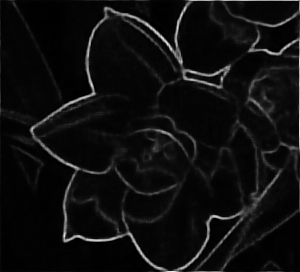}\vspace{\picvspace}
\centering{\scriptsize{Output of the gradient branch}}
\end{minipage}\vspace{\packvspace}

\begin{minipage}[b]{\picwidth\linewidth}
\includegraphics[width=1 \linewidth]{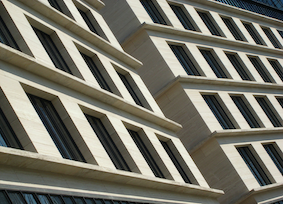}\vspace{\picvspace}
\centering{\scriptsize{HR ('img\_025' from Urban100)}}
\end{minipage}\hspace{\pagehspace}
\begin{minipage}[b]{\picwidth\linewidth}
\includegraphics[width=1 \linewidth]{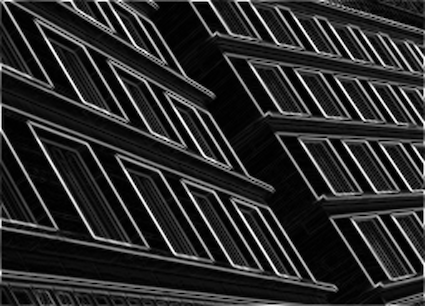}\vspace{\picvspace}
\centering{\scriptsize{LR gradient (Bicubic)}}
\end{minipage}\hspace{\pagehspace}
\begin{minipage}[b]{\picwidth\linewidth}
\includegraphics[width=1 \linewidth]{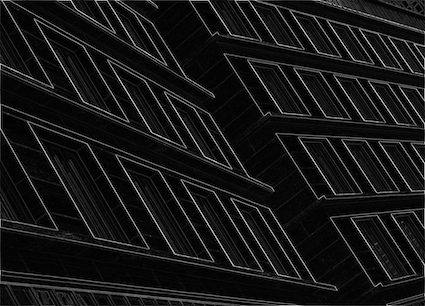}\vspace{\picvspace}
\centering{\scriptsize{HR gradient}}
\end{minipage}\hspace{\pagehspace}
\begin{minipage}[b]{\picwidth\linewidth}
\includegraphics[width=1 \linewidth]{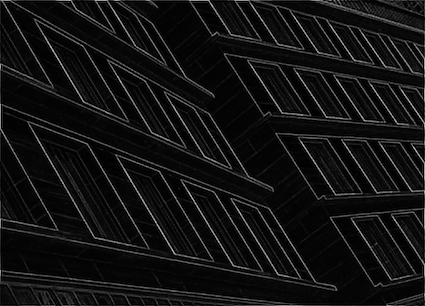}\vspace{\picvspace}
\centering{\scriptsize{Output of the gradient branch}}
\end{minipage}\vspace{\captionvspace}

\centering
\caption{Visualization of gradient maps. }
\label{fig:supp_grad}
\end{figure*}

We display more SR performance comparison with state-of-the-art SR methods including EnhanceNet~\cite{EnhanceNet}, SFTGAN~\cite{SFTGAN}, SRGAN~\cite{ledig2017photo}, ESRGAN~\cite{wang2018esrgan} and NatSR~\cite{soh2019natural}, 
as shown in Figure~\ref{fig:supp_vis1},~\ref{fig:supp_vis2},~\ref{fig:supp_vis3},~\ref{fig:supp_vis4} and~\ref{fig:supp_vis5}.  
The results show our SPSR method performs better than other SR methods in recovering structural-pleasant and photo-realistic images. 
We also visualize the outputs of the gradient branch, as shown in Figure~\ref{fig:supp_grad}. We can see the gradient branch succeeds in converting LR gradient maps to the HR ones.

\end{document}